\pdfoutput=1

\documentclass[11pt,twoside,a4paper,cmspaper,final,collab]{cms-tdr}
\def\svnVersion{398794}\def\cmsCernNoTag{CERN-EP/2016-242}\def\cmsCernDate{\today}\def\cmsMessage{Published in the Journal of High Energy Physics as \href{http://dx.doi.org/10.1007/JHEP04(2017)039}{\doi{10.1007/JHEP04(2017)039.}}}

\begin{document}\cmsNoteHeader{HIN-15-015}

\hyphenation{had-ron-i-za-tion}
\hyphenation{cal-or-i-me-ter}
\hyphenation{de-vices}
\RCS$Revision: 389117 $
\RCS$HeadURL: svn+ssh://svn.cern.ch/reps/tdr2/papers/HIN-15-015/trunk/HIN-15-015.tex $
\RCS$Id: HIN-15-015.tex 389117 2017-02-23 19:25:09Z alverson $
\newlength\cmsFigWidth
\ifthenelse{\boolean{cms@external}}{\setlength\cmsFigWidth{0.49\textwidth}}{\setlength\cmsFigWidth{0.70\textwidth}}
\ifthenelse{\boolean{cms@external}}{\providecommand{\cmsLeft}{top\xspace}}{\providecommand{\cmsLeft}{left\xspace}}
\ifthenelse{\boolean{cms@external}}{\providecommand{\cmsRight}{bottom\xspace}}{\providecommand{\cmsRight}{right\xspace}}
\ifthenelse{\boolean{cms@external}}{\providecommand{\cmsLLeft}{Top\xspace}}{\providecommand{\cmsLLeft}{Left\xspace}}
\ifthenelse{\boolean{cms@external}}{\providecommand{\cmsRRight}{Bottom\xspace}}{\providecommand{\cmsRRight}{Right\xspace}}

\newcommand{\PbPb}{\ensuremath{\mathrm{PbPb}}\xspace}
\newcommand{\pp}{\ensuremath{\Pp\Pp}\xspace}
\newcommand{\raa}{\ensuremath{R_\mathrm{AA}}\xspace}
\newcommand{\rpa}{\ensuremath{R_{\Pp\mathrm{A}}}\xspace}
\providecommand{\sqrtsnn}{\ensuremath{\sqrt{\smash[b]{s_\mathrm{NN}}}}\xspace}
\providecommand{\NA}{\ensuremath{\text{---}\xspace}}

\cmsNoteHeader{HIN-15-015}
\title{Charged-particle nuclear modification factors
in PbPb and pPb collisions at $\sqrtsnn=5.02$\TeV}

\date{\today}

\abstract{The spectra of charged particles produced within the pseudorapidity window
$\abs{\eta}<1$ at $\sqrtsnn=5.02$\TeV are measured using 404\mubinv of PbPb and
27.4\pbinv of pp data collected by the CMS detector at the LHC in 2015. The spectra are
presented over the transverse momentum ranges spanning $0.5<\pt<400$\GeV in pp and
$0.7<\pt<400$\GeV in PbPb collisions. The corresponding nuclear modification factor,
\raa, is measured in bins of collision centrality. The \raa in the
5\% most central collisions shows a maximal suppression by a factor of
7--8 in the \pt region of 6--9\GeV. This dip is
followed by an increase, which continues up to the highest \pt measured, and approaches unity in the vicinity of $\pt=200$\GeV. The \raa is
compared to theoretical predictions and earlier experimental results at lower collision
energies. The newly measured pp spectrum is combined with the pPb spectrum previously
published by the CMS Collaboration to construct the pPb nuclear modification factor,
\rpa, up to 120\GeV. For $\pt>20$\GeV, \rpa exhibits weak momentum
dependence and shows a moderate enhancement above unity.}

\hypersetup{%
pdfauthor={CMS Collaboration},%
pdftitle={Charged-particle nuclear modification factors in PbPb and pPb collisions at sqrt(s[NN)]=5.02 TeV},%
pdfsubject={CMS},%
pdfkeywords={CMS, physics, heavy ions, nuclear modification factor}}

\maketitle

\section{Introduction}
\label{sec:intro}

The charged-particle transverse momentum (\pt) spectrum is an
important tool for studying parton energy loss in the dense QCD medium,
known as the quark gluon plasma (QGP), that is produced in high energy
nucleus-nucleus (AA) collisions~\cite{Bjorken:1982tu,d'Enterria:2009am}.
In such collisions, high-\pt particles,
which originate from parton fragmentation, are sensitive to the amount of
energy loss that the partons experience traversing the medium. By
comparing high-\pt particle yields in AA collisions
to predictions of theoretical models, insight into the
fundamental properties of the QGP can be gained. Over the years, a
number of results have been made available by experiments at SPS~\cite{Aggarwal:2001gn,d'Enterria:2004ig}, at RHIC~\cite{Arsene:2004fa,Back:2004je,Adams:2005dq,Adcox:2004mh},
and at the CERN
LHC~\cite{Abelev:2012hxa,Aad:2015wga,CMS:2012aa}. The modification of
high-\pt particle production is typically quantified using the
ratio of the charged-particle \pt spectrum in AA
collisions to that of pp collisions, scaled by the average number of binary
nucleon-nucleon collisions, $\langle N_\text{coll} \rangle$. This quantity is known as the
nuclear modification factor, \raa, and
can also be formulated as function of \pt as
\begin{equation}
\raa(\pt) = \frac{\rd N^\mathrm{AA}/\rd\pt}{\langle N_\text{coll}\rangle \rd N^{\Pp\Pp}/\rd\pt}=
\frac{\rd N^\mathrm{AA}/\rd\pt}{T_\mathrm{AA}\,\rd\sigma^{\Pp\Pp}/\rd\pt},
\label{eqn:def_raa}
\end{equation}
where $N^\mathrm{AA}$ and $N^{\Pp\Pp}$ are the charged-particle yields in
AA collisions and pp collisions, and
$\sigma^{\Pp\Pp}$ is the charged-particle cross section in pp
collisions. The ratio of $\langle N_\text{coll} \rangle$ with the total
inelastic pp cross section,
defined as $T_\mathrm{AA}$ = $\langle N_\text{coll} \rangle /\sigma^{\Pp\Pp}_\text{inel}$, is known as the
nuclear overlap function and can be calculated from a Glauber model of
the nuclear collision geometry~\cite{Miller:2007ri}.  In this work we adopt natural units, such that $c=1$.

The factor of 5 suppression observed in the \raa of charged
hadrons and neutral
pions at RHIC~\cite{Arsene:2004fa,Adcox:2004mh,Adams:2005dq,Back:2004je}
was an indication of strong medium effects on particle production in the
final state. However, the RHIC measurements were limited to a \pt range below 25\GeV and a collision energy per nucleon
pair,  \sqrtsnn, less than or equal to 200\GeV. The QGP is
expected to have
a size, lifetime, and temperature that are affected by the collision
energy. During the first two PbPb runs, the LHC collaborations measured
the charged-particle \raa at $\sqrtsnn=2.76$\TeV, up to \pt around
50\GeV ({ALICE}~\cite{Abelev:2012hxa}), 100\GeV (CMS~\cite{CMS:2012aa}), and 150\GeV (ATLAS~\cite{Aad:2015wga}).
A suppression by a factor of about 7 was observed in the 5--10\GeV
\pt region \cite{Abelev:2012hxa,Aad:2015wga,CMS:2012aa}. At higher \pt,
the suppression was not as strong, approaching roughly a factor of 2
for particles with \pt in the range of 40--100\GeV. At the end of
2015, in the first heavy ion data-taking period of the Run-2 at the LHC, PbPb
collisions at $\sqrtsnn=5.02$\TeV took place, allowing the study of the suppression of charged particles at a
new collision energy frontier. Proton-proton data at the same
collision energy were also taken, making direct comparison between
particle production in \pp and \PbPb collisions possible.

To gain access to the properties of the QGP,
it is necessary to separate the effects directly related to the hot
partonic QCD system from those referred to as cold nuclear
matter effects. Measurements
in proton-nucleus collisions can be used for this purpose. The CMS
Collaboration has
previously published results for the nuclear modification factor $\rpa^{*}$ using measured charged-particle spectra in pPb collisions at
$\sqrtsnn=5.02$\TeV and a pp reference spectrum constructed by
interpolation from previous measurements at higher and lower
center-of-mass energies~\cite{Khachatryan:2015xaa}. The asterisk in the
notation refers to this usage of an interpolated reference spectrum.
Similarly interpolation-based results are also available
from the ATLAS~\cite{Aad:2016zif} and the ALICE~\cite{ALICE:2012mj} experiments.
With the pp data taken in 2015 at $\sqrt{s} = 5.02$\TeV, the measurement
of the nuclear modification factor, \rpa, using a measured
pp reference spectrum, becomes possible.

In this paper, the spectra of charged particles in the pseudorapidity window $\abs{\eta}<1$ in pp and PbPb collisions at $\sqrtsnn=5.02$\TeV, as well as the nuclear modification factors, \raa and \rpa, are presented.
Throughout this paper, for each collision system, the pseudorapidity is computed in the
center-of-mass frame of the colliding nucleons. The measured \raa is
compared to model calculations, as well as to previous experimental
results at lower collision energies.

\section{The CMS detector and data selection}
\label{sec:samples_triggers}
The central feature of the CMS apparatus is a superconducting solenoid of
6\unit{m} internal diameter, providing an axial magnetic field of 3.8\unit{T}. Within
the solenoid volume are a silicon pixel and strip tracker
covering the range of $\abs{\eta}<2.5$~\cite{Chatrchyan:2008zzk}, a
lead tungstate crystal electromagnetic calorimeter, and a brass and
scintillator hadron calorimeter, each composed of a barrel and two endcap
sections. Hadron forward calorimeters (HF), consisting of steel with embedded quartz fibers, extend the
calorimeter coverage up to $\abs{\eta}<5.2$. Muons are measured in gas-ionization
detectors embedded in the steel flux-return yoke outside the solenoid. A more
detailed description of the CMS detector, together with a definition of the
coordinate system used and the relevant kinematic variables, can be found in
Ref.~\cite{Chatrchyan:2008zzk}.

The measurement of \raa is performed using the 2015 pp and PbPb data
taken at $\sqrtsnn=5.02$\TeV. The pp sample corresponds to an
integrated luminosity of 27.4\pbinv, while the PbPb sample
corresponds to an integrated luminosity of 404\mubinv.  For pp collisions the average pileup (the mean of the Poisson distribution of the
number of collisions per bunch crossing) was approximately 0.9.  For the
measurement of \rpa, 35\nbinv of $\sqrtsnn=5.02$\TeV
pPb data are used.

The collision centrality in PbPb events, \ie the degree of overlap of the two colliding
nuclei, is determined from the total transverse energy, \ET, deposition in both HF
calorimeters. Collision-centrality bins are given in percentage ranges of the total hadronic
cross section, 0--5\% corresponding to the 5\% of collisions with the largest overlap of the
two nuclei. The collision centrality can be related to properties of the PbPb collisions,
such as the total number of binary nucleon-nucleon collisions, $N_\text{coll}$. The
calculation of these properties is based on a Glauber model of the incoming
nuclei and their constituent nucleons~\cite{Miller:2007ri,Alver:2008zza}, as well as studies of bin-to-bin smearing, which is
evaluated by examining the effects of finite resolution on fully simulated and reconstructed
events~\cite{Chatrchyan:2011sx}.  The calculated average $N_\text{coll}$ and $T_\mathrm{AA}$
values corresponding to the centrality ranges used, along with their systematic
uncertainties, are listed in Table~\ref{tab:TAA}. The $\sigma^{\Pp\Pp}_\text{inel}$ utilized in
the Glauber calculation is $70\pm5$\unit{mb}~\cite{Agashe:2014kda}.  The
nuclear radius and skin depth are $6.62 \pm 0.06$\unit{fm} and $0.546 \pm
0.010$\unit{fm}, respectively, and a minimal distance between nucleons of
$0.04 \pm 0.04$\unit{fm} is imposed~\cite{DEVRIES1987495}.  In this paper, only
$T_\mathrm{AA}$ is used in the calculation of \raa, as given by the last
formula in Eq.~(\ref{eqn:def_raa}).

\begin{table}[tbh]
\centering
\topcaption{The values of $\langle N_\text{coll} \rangle$ and $T_\mathrm{AA}$ and their
uncertainties in $\sqrtsnn=5.02$\TeV PbPb collisions for the
centrality ranges used in this paper.}
\newcolumntype{x}{D{,}{\text{--}}{2.3}}
\newcolumntype{z}{D{,}{}{5.5}}
\setlength\extrarowheight{1.5 pt}
\begin{tabular}{xzz}
 \hline
  \multicolumn{1}{c}{Centrality} &  \multicolumn{1}{c}{$\langle N_\text{coll} \rangle$} &  \multicolumn{1}{c}{$T_\mathrm{AA}$ [mb$^{-1}$]} \\
 \hline
0,5\% & 1820,{^{+130}_{-140}} & 26.0,{^{+0.5}_{-0.8}} \\[1.5pt]
5,10\% & 1430,{^{+100}_{-110}} & 20.5,{^{+0.4}_{-0.6}} \\[1.5pt]
10,30\% & 805,{^{+55}_{-58}} & 11.5,{^{+0.3}_{-0.4}} \\[1.5pt]
30,50\% & 267,{^{+20}_{-20}} & 3.82,{^{+0.21}_{-0.21}} \\[1.5pt]
50,70\% & 65.4,{^{+7.0}_{-6.6}} & 0.934,{^{+0.096}_{-0.089}} \\[1.5pt]
70,90\% & 10.7,{^{+1.7}_{-1.5}} & 0.152,{^{+0.024}_{-0.021}} \\[1.5pt]
0,10\% & 1630,{^{+120}_{-120}} & 23.2,{^{+0.4}_{-0.7}} \\[1.5pt]
0,100\% & 393,{^{+27}_{-28}} & 5.61,{^{+0.16}_{-0.19}} \\[1.5pt]
 \hline
\end{tabular}
\label{tab:TAA}
\end{table}

The CMS online event selection employs a hardware-based level-1 trigger
(L1) and a software-based high-level trigger (HLT). Minimum-bias $\pp$ and
\PbPb\ collisions were selected using an HF-based L1 trigger requiring
signals above threshold in either one (\pp) or both (\PbPb) sides of HF
calorimeters. These data were utilized to access the low-\pt kinematic
region of charged particles. In order to extend the \pt reach of the
results reported in this paper, events selected by jet triggers were
used.  High-\pt track triggers were also employed, but only as a
cross-check of the result obtained with jet triggers.

\begin{table}[tbh]
\centering
\topcaption{Summary of the \ET and \pt thresholds of the various
L1 and HLT triggers
used in the analysis for the two colliding systems. Please refer to the
text about the exact meaning of the thresholds. Only the
highest-threshold triggers collected data
unprescaled. The MB symbol refers to seeding by a minimum-bias trigger.}
\begin{tabular}{lll}
 \hline
  Collision system/trigger &  \multicolumn{1}{c}{L1 thresholds [\GeVns{}]} &  \multicolumn{1}{c}{HLT thresholds [\GeVns{}]} \\
 \hline
  pp & & \\
  \rule{10pt}{0pt}Jet triggers & 28, 40, 48 & 40, 60, 80 \\
  \rule{10pt}{0pt}Track triggers & MB, 28, 40, 48 & 12, 24, 34, 45, 53 \\
  PbPb & & \\
  \rule{10pt}{0pt}Jet triggers & 28, 44, 56 & 40, 60, 80, 100 \\
  \rule{10pt}{0pt}Track triggers & MB, 16, 24 & 12, 18, 24, 34 \\
 \hline
\end{tabular}
\label{tab:triggers}
\end{table}

At the L1 stage, the jet-triggered events in pp and PbPb collisions were
selected by requiring the presence of L1-recon\-struct\-ed jets above
various \ET thresholds, listed in Table~\ref{tab:triggers}. While
the lower-threshold triggers had to be
prescaled because of the high instantaneous luminosity of the LHC, the
highest threshold trigger was always unprescaled. In
PbPb collisions, the L1 jet trigger algorithms performed an online event-by-event
underlying-event subtraction, estimating the energy of the underlying
event by averaging the deposited calorimeter \ET in rings of
azimuthal angle ($\phi$, in radians) as a function of $\eta$, for each event
separately. Events triggered by high-\pt tracks in
pp collisions were selected by the same L1 jet triggers as described
above. In PbPb collisions, a special algorithm based on the \ET of the
highest-\ET underlying-event subtracted calorimeter trigger
region ($\Delta\eta$, $\Delta\phi=0.348$) in the central
($\abs{\eta}<1.044$) detector area was employed. The presence of a high-\pt
track is better
correlated with the presence of a high-\ET trigger region than
with the presence of a multiregion-wide L1 jet. Therefore, seeding the
high-\pt track triggers with the former algorithm leads to a lower
overall L1 trigger rate. This was an important consideration in PbPb
collisions, while it had much less importance in pp ones.
Both the jet and the track triggers had variants selecting only PbPb collision events of specific centralities. This was made
possible by an L1 algorithm, which estimated the collision centrality
based on the sum of the \ET deposited in the HF calorimeter
regions. The measurement of PbPb spectra reported in this paper makes
use of such triggers to increase the number of events in peripheral
centrality bins.

At the HLT, online versions of the pp and PbPb offline calorimeter jet
and track reconstruction algorithms were run. In pp collisions, events
selected by high-level jet triggers contain calorimeter clusters which
are above various \pt values (Table~\ref{tab:triggers}) in the
$\abs{\eta}<5.1$ region.  Such clusters were produced with the
anti-\kt algorithm~\cite{Cacciari:2008gp, Cacciari:2011ma} of
distance parameter R=0.4, and were
corrected to establish a relative uniform calorimeter response in $\eta$
and a calibrated absolute response in \pt.  In this configuration, the
80\GeV threshold trigger was unprescaled. In PbPb collisions, the R=0.4
anti-\kt calorimeter jets were clustered and corrected after
the energy due to the heavy-ion underlying event was subtracted in an
$\eta$-dependent way~\cite{Kodolova:2007hd}. Triggers with thresholds on
the jet energy from 40 to 100\GeV were employed. The independent
high-\pt track triggers looked for a track in the $\abs{\eta}<2.4$ (pp) and
$\abs{\eta}<1.05$ (PbPb) regions above different \pt thresholds, listed in Table~\ref{tab:triggers}.

Events selected for offline analysis are required to pass a set of
selection criteria designed to reject events from background processes
(beam-gas collisions and beam scraping events). Events are required to have
at least one reconstructed primary interaction vertex with at least two
associated tracks. In pp collisions, the events are also required to
have at least 25\% of the tracks passing a tight track-quality selection
requirement~\cite{1748-0221-9-10-P10009}. In PbPb collisions, the
shapes of the clusters in the pixel detector are required to be
compatible with those expected from particles produced by a PbPb
collision. The PbPb collision event is also required to have at least
three towers in each of the HF detectors with energy deposits of more
than 3\GeV per tower.

\section{Track reconstruction and corrections}
\label{sec:trking}

The distributions reported in this paper are for primary charged
particles.  Primary charged particles are required to have a mean proper
lifetime greater than 1 \rm{cm}.  The daughters of secondary decays are
considered primary only if the mother particle had a mean proper
lifetime less than 1 \rm{cm}.  Additionally, charged particles resulting
from interactions with detector material are not considered primary
particles.

The track reconstruction used in pp collisions for this study is
described in Ref.~\cite{1748-0221-9-10-P10009}.  In PbPb collisions,
minor modifications are made to the pp algorithm in order to accommodate
the much larger track multiplicities.  Only tracks in the range
$\abs{\eta}<1$ are used.  Tracks are required to have a relative \pt
uncertainty of less than 10\% in PbPb collisions and 30\% in pp
collisions. In PbPb collisions, tracks must also have at least 11 hits and satisfy a stringent fit
quality requirement, specifically that the $\chi^2$, divided by both the number of degrees of freedom and the number of tracker layers hit, be less than 0.15. To decrease the
likelihood of counting nonprimary charged particles originating from
secondary decay products, a selection requirement of less than
3 standard deviations is applied on the significance of the distance of
closest approach to at
least one primary vertex in the event, for both collision systems.
Finally, a selection based on the relationship of a track to calorimeter
energy deposits along its trajectory is applied in order to
curtail the contribution of misreconstructed tracks with very high \pt.
Tracks with $\pt>20$\GeV are required to have an associated energy
deposit~\cite{CMS-PAS-PFT-10-002} of at least half their momentum in the CMS calorimeters. This
requirement was determined by comparing the distributions of the
associated deposits for genuine and misreconstructed tracks in simulated
events to tracks reconstructed in real data. The efficiency of the
calorimeter-matching requirement is 98\%\,(95\%) in PbPb\,(pp) data for
tracks selected for analysis by the previously mentioned other track
selection criteria.

To correct for inefficiencies associated with the track reconstruction
algorithms, simulated Monte Carlo (MC) samples are used. For pp collision data, these are
generated with \PYTHIA 8.209~\cite{Sjostrand:2007gs}
tune CUETP8M1~\cite{Khachatryan:2015pea} minimum-bias, as well as QCD dijet samples
binned in the transverse momentum of the hard scattering, $\hat{p}_\mathrm{
T}$. For PbPb collision
data, \textsc{hydjet} 1.9~\cite{Lokhtin:2005px} minimum-bias events and \textsc{hydjet}-embedded \PYTHIA QCD dijet events are used. In the embedding
procedure, a high-$\hat{p}_\mathrm{T}$ \PYTHIA event is combined with a
minimum-bias \textsc{hydjet} event with the same vertex location.  The
combined event is then used as input to the full simulation of the CMS
detector response.

In general the tracking efficiency, defined as the fraction of primary
charged particles successfully reconstructed, is non-unitary due to
algorithmic inefficiencies and detector acceptance effects.
Furthermore, misreconstruction, where a track not corresponding to
any charged particle is errantly reconstructed, can inject extra tracks
into the analysis.  Finally, tracks corresponding to products of
secondary interactions or decays, which still pass all track selection
criteria and are therefore selected for analysis, must also be taken
into account.  Corrections for these effects are applied on a
track-by-track basis, and take into consideration
the properties of each track: \pt, $\eta$, $\phi$, and
radial distance of the track from the closest jet axis.  The functional
dependence of the corrections is assumed to factorize into the product
of four single-variable functions in separate classes of track
kinematics properties.  This factorization is only approximate
because of correlations between the variables. These correlations are
accounted for in a systematic uncertainty.  The tracking efficiency in
pp is between 80 and 90\% for
most of the \pt range studied, except for $\pt>150$\GeV,
where it decreases to 70\%.  The pp track misreconstruction rate and secondary rate are found
to be less than 3\% and 1\%, respectively, in each \pt bin examined.
Owing to the
dependence of the tracking efficiency on detector occupancy, the event
centrality is also taken into account in the correction procedure for PbPb collisions.
Additionally, to account for the slightly different $\chi^{2}/\mathrm{dof}$
in data and simulated events, a track-by-track reweighting is applied to
the simulation during this calculation. The efficiency of the PbPb track
reconstruction algorithm and track selection criteria for minimum-bias
events is approximately 40\% at 0.7\GeV. It then increases rapidly to
around 65\%
at 1\GeV, where it reaches a plateau. It starts to decrease from \pt
values of around 100\GeV until it reaches about 50\% at 400\GeV. This
efficiency is also centrality dependent; the \pt-inclusive value is approximately 60\% for
central events and 75\% for peripheral events. In general, the PbPb
misreconstruction and secondary rates are very small because of the
strict selection criteria applied to the tracks. The misreconstruction rate does increase
at low track \pt and also slightly at very high \pt, to around 1.5\%.
Below 1\GeV it increases to 10\% for the most
central events.  These numbers are in line with the expected tracking performance based on previous studies of similar tracking algorithms in pp collisions at $\sqrt{s} = 7$\TeV \cite{1748-0221-9-10-P10009} and PbPb collisions at $\sqrtsnn=2.76$\TeV~\cite{MPT_Khachatryan:2015lha}.

Particles of different species have different track reconstruction
and selection efficiencies at the same \pt. As different MC event
generators model the relative fractions of the particle species
differently, the computed tracking efficiencies for inclusive primary
charged particles depend on which MC generator is used to evaluate the
correction. Notably, the reconstruction efficiency for primary charged
strange baryons is very low, as they decay before leaving a sufficient
number of tracker hits for direct reconstruction. In
this measurement, the species-dependent track reconstruction
efficiencies are first calculated and then weighted with the corresponding particle fractions
produced by \PYTHIA 8, tune CUETP8M1 and \textsc{epos}~\cite{Werner:2005jf}, tune LHC~\cite{Pierog:2013ria}.  \PYTHIA~is expected to underpredict the fraction of strange baryons present in PbPb collisions, while \textsc{epos} overpredicts strange baryon production in central collisions at lower collision center-of-mass energies~\cite{ABELEV:2013zaa}.  Therefore we choose a working point between these two models by averaging the two sets of correction factors.

The \pt resolution of selected tracks in both pp and PbPb collisions
remains below 2\% up to 100\GeV.  For higher \pt it
starts to increase, reaching about 6\% at 400\GeV. The resulting change
in the measured charged-particle yields introduced by the track
resolution is found to be less than 1\%. A correction is not made for
this distortion, but rather the distortion is accounted for in the
systematic uncertainty.

The distortion of the shape of the pp \pt distribution due to the event
selection requirements is calculated by evaluating the
efficiency of the selection in ``zero bias'' data. Zero bias data were
selected solely based on whether there were filled bunches in both beams
crossing each other in the CMS interaction region. Therefore, the zero
bias data set provides an unbiased sample to study the efficiency of the
minimum-bias trigger and of the offline event selection.  As a result of
this study, a correction is applied for a small (less than 1\%)
distortion of the very low-\pt spectrum due to valid events failing to pass the
event selection.  For the PbPb sample, the event selection is fully
efficient from 0 to 90\% event centrality classes.  For quantities
inclusive in
centrality, the event selection efficiency of $99 \pm2\%$ is corrected
for. (Selection efficiencies higher than 100\% are possible, reflecting the
presence of ultra-peripheral collisions in the selected event
sample.)

\section{Combination of data from different triggers}
\label{sec:eventClass}
To obtain the inclusive charged-particle spectra up to a few hundred \GeV
of transverse momenta, data recorded by the minimum-bias and jet
triggers are combined. The procedure is outlined in
Refs.~\cite{CMS:2012aa, Khachatryan:2015xaa}.

The event-weighting factors corresponding to the various triggers are
computed by counting the number of events that contain a leading jet
(defined as the jet with the highest \pt in the event) in the range of
$\abs{\eta} < 2$ with \pt values in regions not affected by trigger
thresholds. In these regions, the trigger efficiency of the
higher-threshold trigger is constant relative to that of the
lower-threshold trigger. The ratio of the number of such events in the
two triggered sets of data is
used as a weighting factor. For example, the region above which the jet
trigger with a \pt threshold of 40\GeV has constant efficiency is
determined by comparing the \pt distribution of the leading jets to that
of the minimum-bias data. Similarly, the constant efficiency region of
the 60\GeV jet trigger is determined by comparison to the 40\GeV jet
trigger, etc.

\begin{figure}[tbh]
  \centering
    \includegraphics[width=0.49\textwidth]{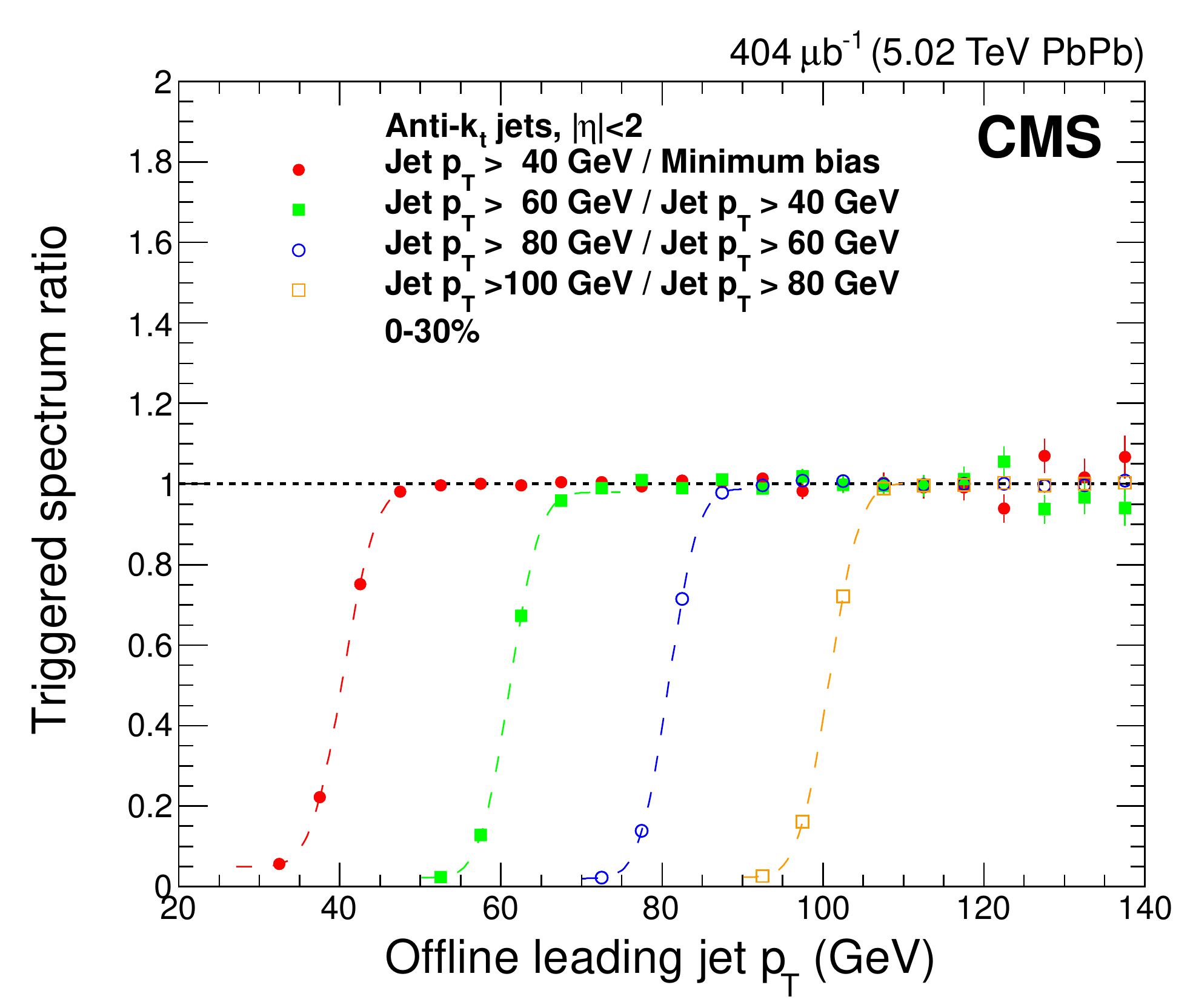}
    \includegraphics[width=0.49\textwidth]{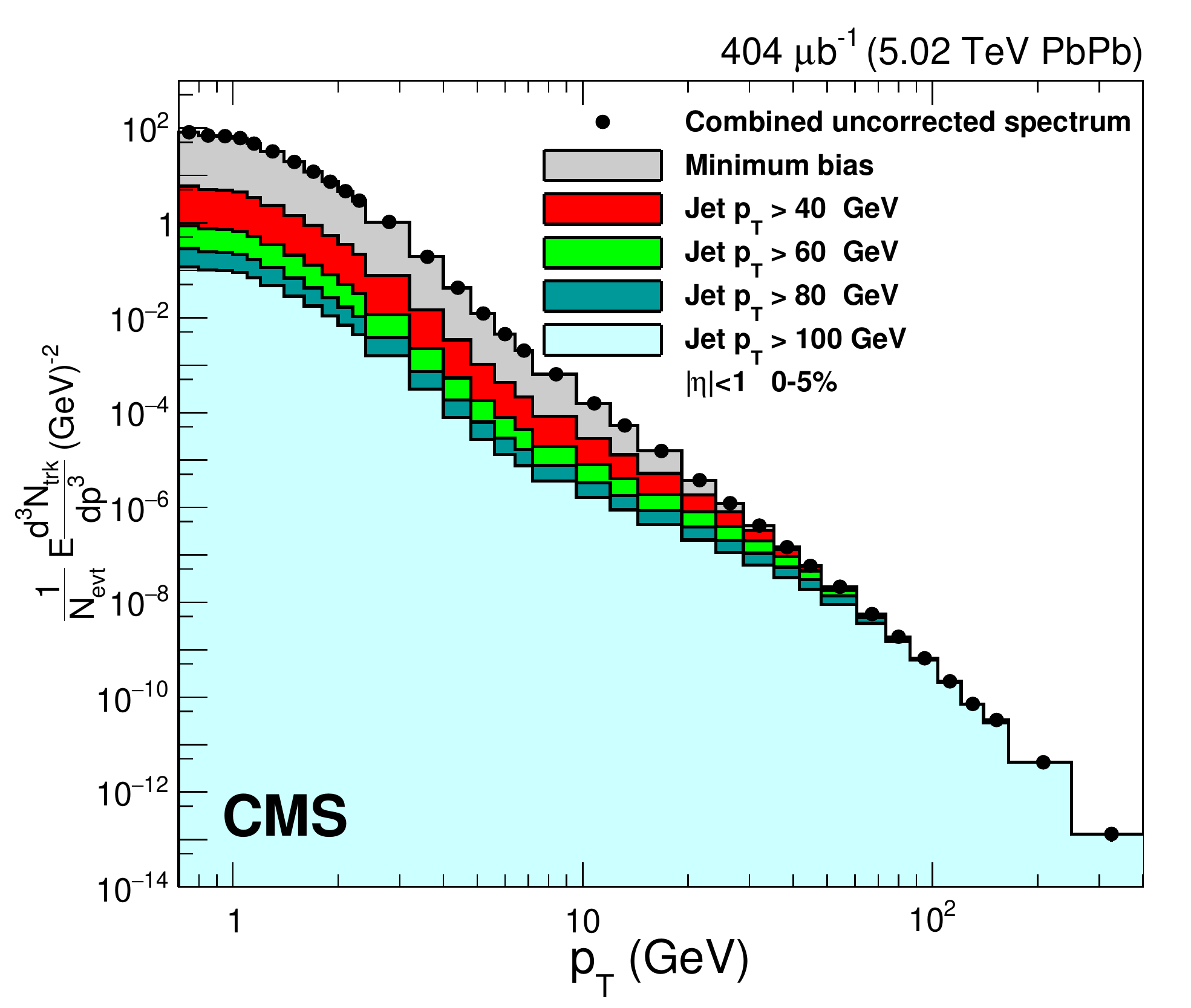}
    \caption{{\cmsLLeft:} ratio of the leading jet \pt distributions in PbPb collisions in the
0--30\% centrality range from various triggers, after the data have
been normalized to one another. Lines have been added to guide the eye.
{\cmsRRight:} contributions from the
various jet triggers (colored
histograms) to the combined, but otherwise uncorrected, track
spectrum (black markers) in the 0--5\% centrality range in PbPb
collisions. The statistical
uncertainties are smaller than the size of the data markers.}
    \label{fig:combJetTriggers_PbPb_ratios}
\end{figure}

To determine the inclusive particle spectrum, events are first uniquely classified
into leading jet \pt classes. The pp spectra are constructed by taking events from the
minimum-bias, 40\GeV jet, 60\GeV jet, 80\GeV jet, and 100\GeV jet triggers, for each respective class. The
particle spectra are evaluated in each class separately, and then combined using the
normalization factors described in the previous paragraph. The procedure outlined
above is verified by constructing a charged-particle spectrum
from an alternative combination of event samples triggered by high-\pt
track triggers. The final spectra are found to be consistent with each
other. In PbPb collisions, the overall normalization of the combined
spectrum is performed using the number of minimum-bias events in the
appropriate centrality range. In pp collisions, the normalization is set
by the integrated luminosity.

The ratio of the normalized distribution of the leading jet \pt from minimum-bias and
from various jet-triggered data in PbPb collisions in the 0--30\% centrality range can
be seen in the left panel of Fig.~\ref{fig:combJetTriggers_PbPb_ratios}. The
constant-efficiency regions are selected to be above \pt of 60, 80, 100, and 120\GeV
for the triggers having a threshold of 40, 60, 80, and 100\GeV, respectively.
The contribution from each of the data sets selected by the different jet
trigger thresholds to the combined, but otherwise uncorrected, track spectrum
 in the 0--5\% centrality range can be seen in the right panel of Fig.~\ref{fig:combJetTriggers_PbPb_ratios}.
The combined spectrum includes contributions from each jet trigger threshold
data set at each charged-particle \pt bin, although the relative
contributions of the different data sets naturally vary strongly as a
function of \pt.

The scheme outlined above is slightly modified for the
combination of the spectra using events from the 0--30\% centrality
range. In that range, due to the large minimum-bias data set and
the absence of the peripheral-specific jet triggers
(see Section~\ref{sec:samples_triggers}), the minimum-bias data provide
higher statistical power than the data triggered with the 40\GeV jet trigger.
Thus, the data from this jet trigger path are not used,
and the minimum-bias sample is combined with the higher-threshold
jet-triggered sample. The 40\GeV jet trigger is shown in
Fig.~\ref{fig:combJetTriggers_PbPb_ratios} for illustration.

\section{Systematic uncertainties}
\label{sec:systUnc}
The systematic uncertainties influencing the measurement of the spectra
of charged particles in pp and PbPb collisions as well as the \raa are
presented in Table~\ref{tab:systUncerts}. The ranges quoted cover both
the \pt and the centrality dependence of the uncertainties. In the
following, each source of systematic uncertainty is discussed
separately, including a discussion on the cancellation of the spectra
uncertainties in \raa.

\begin{itemize}

\item { Particle species composition.  }
As described in Section~\ref{sec:trking}, the tracking corrections used
in the analysis correspond to a particle species composition that lies
halfway between that from \PYTHIA~8, tune CUETP8M1 and \textsc{epos},
tune LHC. We assign the difference between these corrections and the
corrections given by the \PYTHIA~8 or the \textsc{epos} particle
compositions as a systematic uncertainty in the pp and PbPb spectra. The
systematic uncertainty has a strong \pt dependence, directly related to
how much the two models differ at a given \pt. Below a \pt of around
1.5\GeV, the uncertainty is 1\% both in pp and PbPb data. For higher
\pt, the uncertainty increases rapidly with \pt, reaching a value of
about 8\% (pp) and 13.5\% (PbPb in the 0--5\% centrality range) at
3\GeV, followed by a steady decrease to 1\% at and above 10\GeV. The
uncertainties are evaluated in bins of centrality, resulting in higher
uncertainties for more central events. For \raa, the conservative
assumption of no cancellation of this uncertainty is made, resulting in
uncertainty values between 1.5 and 15.5\%.

\item { MC/data tracking efficiency difference.     }
The difference in the track reconstruction efficiency in pp data and pp
simulation was studied by comparing the relative fraction of
reconstructed $\PDast$ mesons in the
$\PDast\to \PD\PGp\to \PK\PGp\PGp$ and
$\PDast\to \PD\PGp\to \PK\PGp\PGp\PGp\PGp$
decay channels in simulated and data events, following
Ref.~\cite{TRK-10-002}. Additional comparisons were made between track
quality variables before track selections in both pp and PbPb data and
simulation.  Based on these two studies, \pt-independent uncertainties
of 4\% (pp) and 5\% (PbPb) are assigned.

To study the potential cancellation of the pp and PbPb uncertainties in
\raa, an examination of the relative difference between pp and PbPb of MC/data tracking
efficiency discrepancies is performed.  First, the ratio of
the uncorrected track spectra in data in the 30--100\% centrality bin
is computed using the pp and the PbPb reconstruction algorithms.
The same ratio is also evaluated using MC
events as inputs. Finally, the ratio of the previously-computed MC
and data ratios is constructed. Assuming that the misreconstruction rate
in data and MC is the same, this double ratio is proportional to the
relative MC/data tracking efficiency difference between pp and PbPb.  Small differences between data and MC, which break the assumption on the misreconstruction rate, are accounted for with the ``fraction of misreconstructed tracks" systematic uncertainty discussed later in this section.
Based on this study, an uncertainty ranging from 2\% (70--90\%
centrality bin) to 6.5\% (0--30\% centrality bins) is assigned to the
\raa\ measurement.

\begin{table}[tbh]
\centering
\topcaption{Systematic uncertainties associated with the measurement of the
charged-particle spectra and \raa using $\sqrtsnn=5.02$\TeV pp
and PbPb collision data. The ranges quoted cover both the \pt and the
centrality dependence of the uncertainties. The combined uncertainty
in \raa does not include the integrated luminosity and the
$T_\mathrm{AA}$ uncertainties.}
\begin{tabular}{lccc}
 \hline
  Sources & \multicolumn{3}{c}{Uncertainty [\%]} \\\cline{2-4}
  & pp & PbPb & \raa \\
 \hline
  Particle species composition & 1--8 & 1.0--13.5 & 1.5--15.5 \\
  MC/data tracking efficiency difference & 4 & 4--5 & 2.0--6.5 \\
  Tracking correction procedure & 1 & 1--4 & 1.5--4.0 \\
  PbPb track selection &\NA& 4 & 4 \\
  Pileup & 3 & $<$1 & 3 \\
  Fraction of misreconstructed tracks & $<$3 & $<$1.5 & $<$3 \\
  Trigger combination & $<$1 & 1 & 1 \\
  Momentum resolution & 1 & 1 & 1 \\
  Event selection correction & $<$1 &\NA& $<$1 \\
 \hline
  Combined uncertainty & 7--10 & 7--15 & 7.0--17.5 \\
 \hline
  Glauber model uncertainty ($T_\mathrm{AA}$) &\NA&\NA& 1.8--16.1 \\
  Integrated luminosity & 2.3 &\NA& 2.3 \\
 \hline
\end{tabular}
\label{tab:systUncerts}
\end{table}

\item { Tracking correction procedure.     }
The accuracy of the tracking correction procedure is tested in simulated
events by comparing the fully corrected track spectrum to the spectrum of
simulated particles. In such comparisons, differences smaller than 1\% (pp)
and 3\% (PbPb) are observed. The main source of the differences is the
fact that the tracking efficiency only approximately factorizes into
single-variable functions of track \pt, track
$\eta$ and $\phi$, event centrality, and radial distance of the tracks from
jets in the bins of track \pt and event centrality used for the
calculation of the tracking correction factors. Such differences in the tracking
corrections are one of the two sources of systematic uncertainty in the
derivation of tracking correction factors considered in this analysis.
The second source of systematic uncertainty is related to only having a
limited number of simulated events to determine the correction factors.
While this uncertainty for
pp collisions is negligible, for PbPb collisions it can reach 3\% and is accounted for in a \pt and centrality-dependent way.
No cancellation of the tracking correction uncertainties in pp and PbPb
collisions is assumed in the computation of \raa.

\item { PbPb track selection.     }
The track selection criteria are stricter in PbPb than in pp collisions.
Selecting on more track quality variables naturally introduces a larger
dependence on the underlying MC/data (dis)agreement for the track
quality variables in question. To study the effect of such
disagreements, the reconstruction of charged-particle spectra was
repeated using looser track selection criteria. Based on the
differences observed in the
fully corrected spectra, an uncertainty of 4\% is assigned for the PbPb
spectra, as well as in \raa.

\item { Pileup.     }
In this analysis, tracks compatible with any of the primary vertices are
selected. To assess the
possible effect of pileup on the particle spectrum, the spectrum was recomputed
using only single-vertex collision events. Based on the differences
observed in the shape of the spectra, a systematic uncertainty of 3\%
is evaluated. For PbPb collisions, the much smaller pileup is found to have
a negligible effect on the reported charged-particle spectra.
Consequently, the 3\% uncertainty in the pp spectrum is propagated to
\raa.

\item {Fraction of misreconstructed tracks.     } The fraction of
misreconstructed tracks is computed from simulated events. To account
for possible differences in the misreconstruction fraction between
simulated and data events, the total amount of the corrections, less
than 3\% in pp and less than 1.5\% in PbPb collisions, is assigned as a
systematic uncertainty in the charged-particle spectra in a
\pt-dependent fashion.  These uncertainties are conservatively assumed to not cancel for the calculation of the uncertainty in \raa.

\item {Trigger combination.     }
The method of combining the different triggers used in this analysis
relies on the calculation of overlaps in the leading jet spectra between
the different triggers. The calculated trigger weights are subject to
statistical fluctuations due to a statistically limited data sample. To
assess the
corresponding uncertainty in \raa, the uncertainties on the trigger
weights associated to each trigger path are weighted according to the
fraction of the particle spectrum that the trigger contributes in a
given \pt bin. The overall uncertainty is found to range from
negligible to 1\%. The uncertainty is highest for peripheral events and
increases with \pt.

\item {Momentum resolution.     }
The variation of the yield of charged particles in any given \pt bin due
to the finite resolution of the track reconstruction is evaluated using
simulated events. The yields are found to only change by around 1\%
both in pp and PbPb collisions. For \raa, the same 1\% systematic
uncertainty is conservatively assigned.

\item {Event selection correction.     }
The bias resulting from the event selection conditions on the shape of
the pp spectrum and \raa distributions is corrected by a procedure,
which directly evaluates the event selection efficiency based
on zero-bias data alone (see Section~\ref{sec:trking}). To
estimate the corresponding systematic uncertainty, the event selection
correction is also evaluated using simulated events. The
charged-particle \pt distribution in pp and the \raa distribution,
reconstructed with the MC-based alternative event selection correction,
are found to differ by less than 1\% from the main result.  For
centrality-inclusive PbPb quantities, an uncertainty due to event
selection is combined with the $T_\mathrm{AA}$ uncertainty.

\item {Glauber model uncertainty.     }
The systematic uncertainty in the Glauber model normalization factor
($T_\mathrm{AA}$) ranges from 1.8\% (in the 0--5\% centrality bin) to 16.1\%
(in the 70--90\% centrality bin). The uncertainties in the $T_\mathrm{AA}$
values are derived from propagating the uncertainties in the event
selection efficiency, and in the nuclear radius, skin depth, and minimum
distance between nucleons in the Pb nucleus~\cite{DEVRIES1987495}
parameters of the Glauber model.

\item {Integrated luminosity.     }
The uncertainty in the integrated luminosity for pp collisions is 2.3\%.
For the PbPb analysis, no luminosity information is used as per-event
yields are measured.

\end{itemize}

\section{Results}
\label{sec:resultSec}
\begin{figure}[tbh]
  \centering
    \includegraphics[width=\cmsFigWidth]{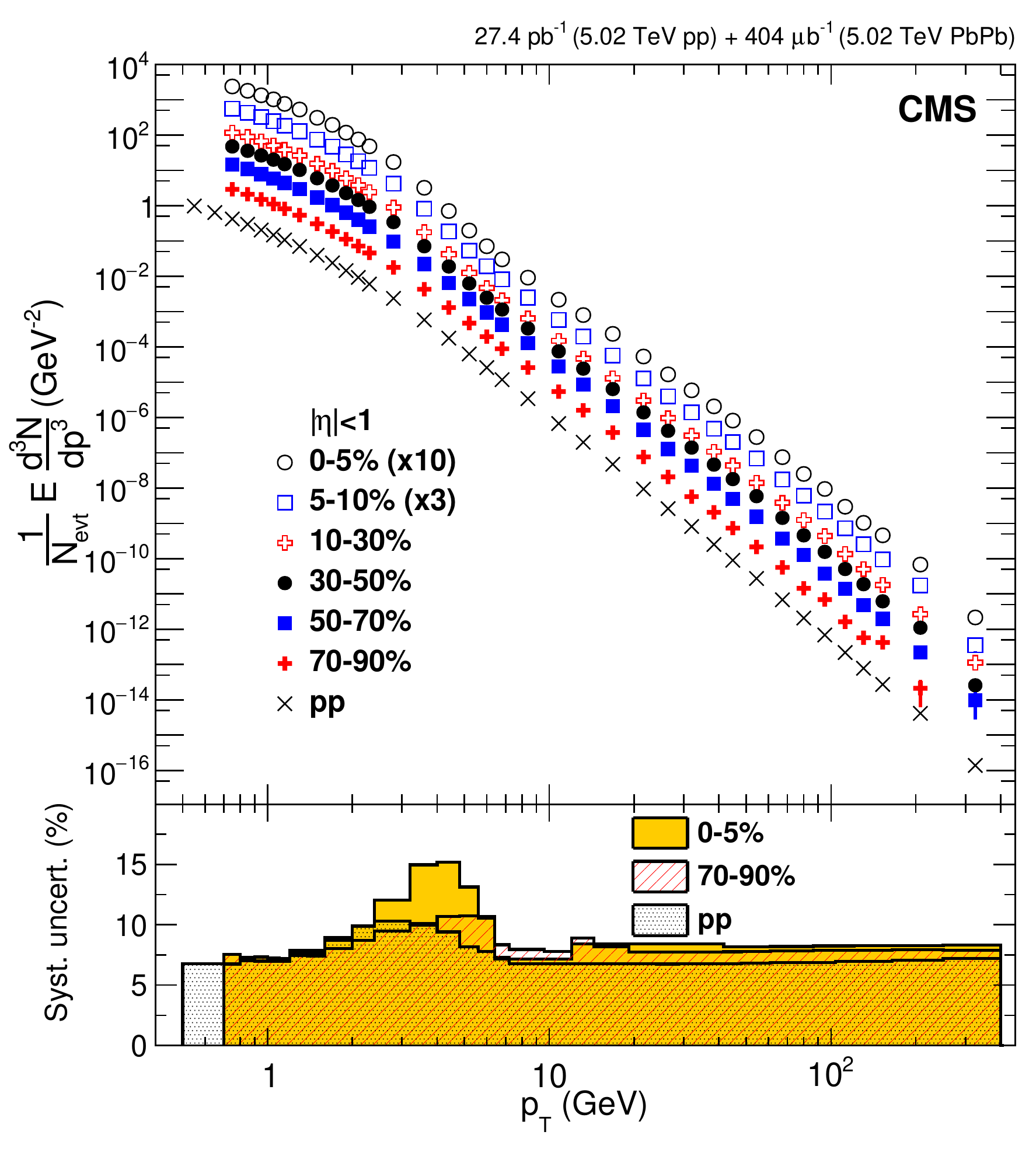}
  \caption{({Top panel}) Charged-particle per-event yields measured in various
PbPb centrality classes, as well as in pp data.  A factor of 70\unit{mb}
is used to scale the pp spectrum from a differential cross section to a
per-event yield for direct comparison.  The statistical uncertainties are smaller than the size of the markers for most points. ({Bottom panel}) Systematic uncertainties
as a function of \pt for representative data sets. The pp uncertainty
contains a 2.3\% fully correlated uncertainty in the pp integrated
luminosity.}
  \label{fig:SpectraResult}
\end{figure}

\begin{figure*}[htb!]
  \centering
    \includegraphics[width=0.45\textwidth]{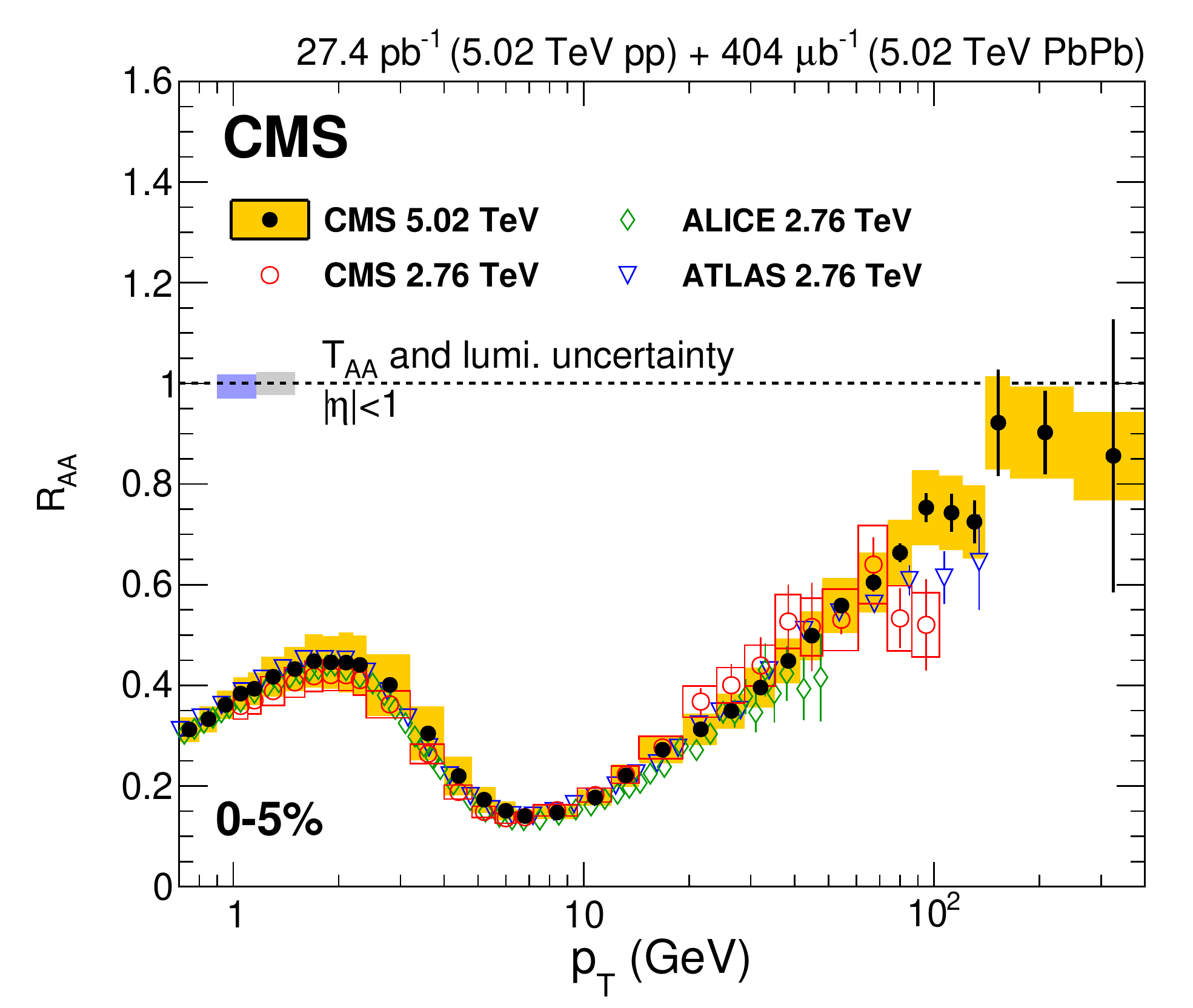}
    \includegraphics[width=0.45\textwidth]{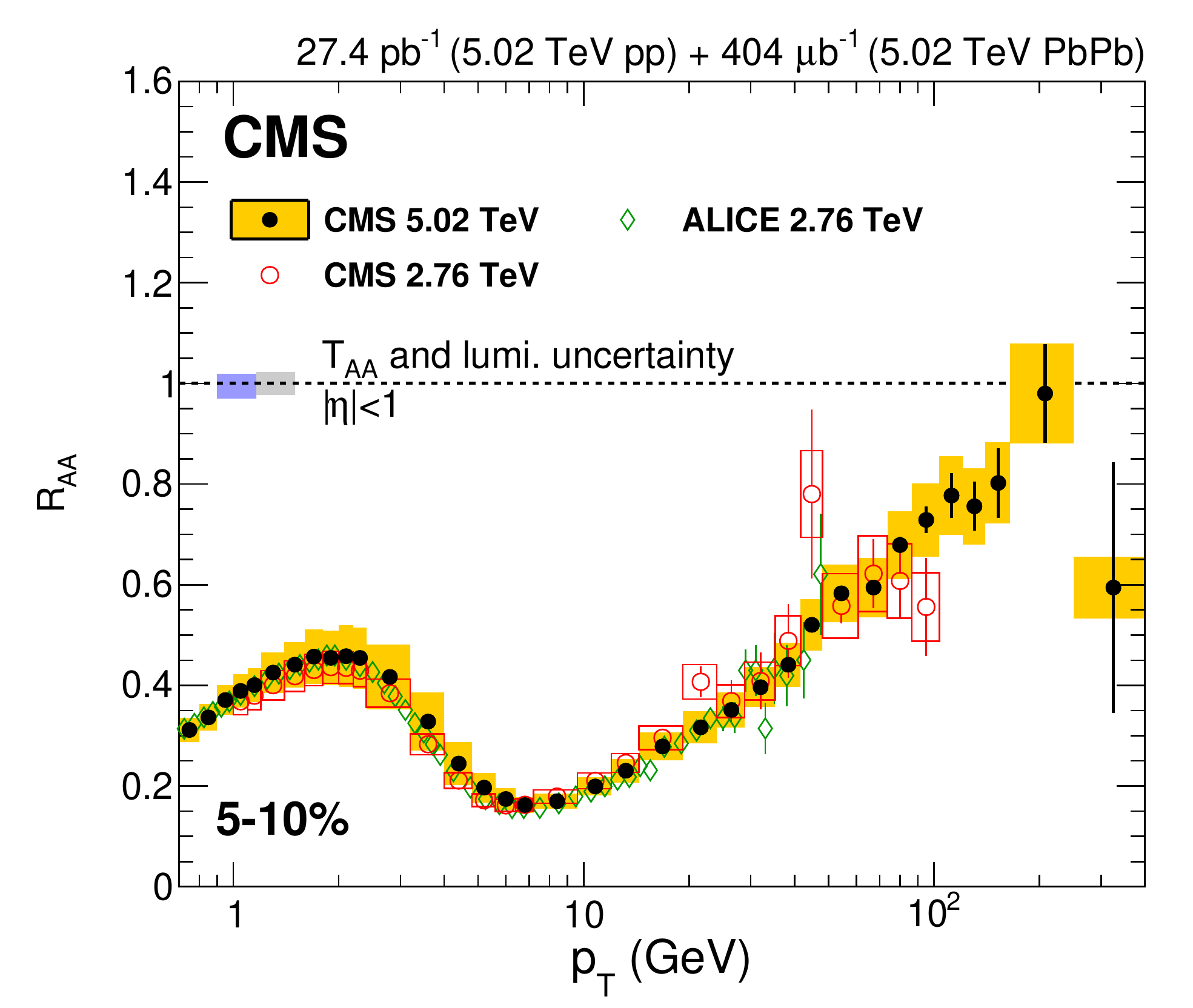}
    \includegraphics[width=0.45\textwidth]{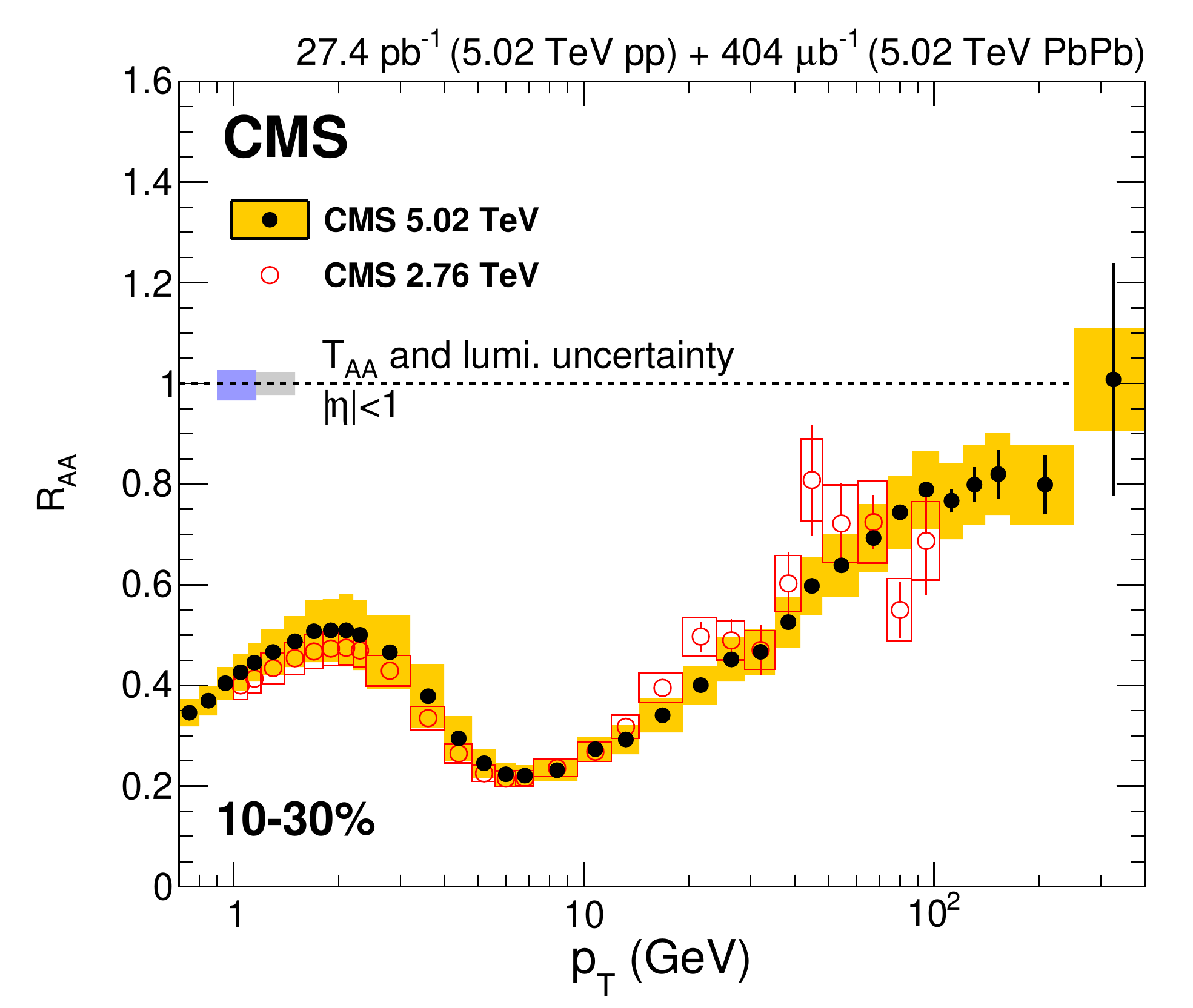}
    \includegraphics[width=0.45\textwidth]{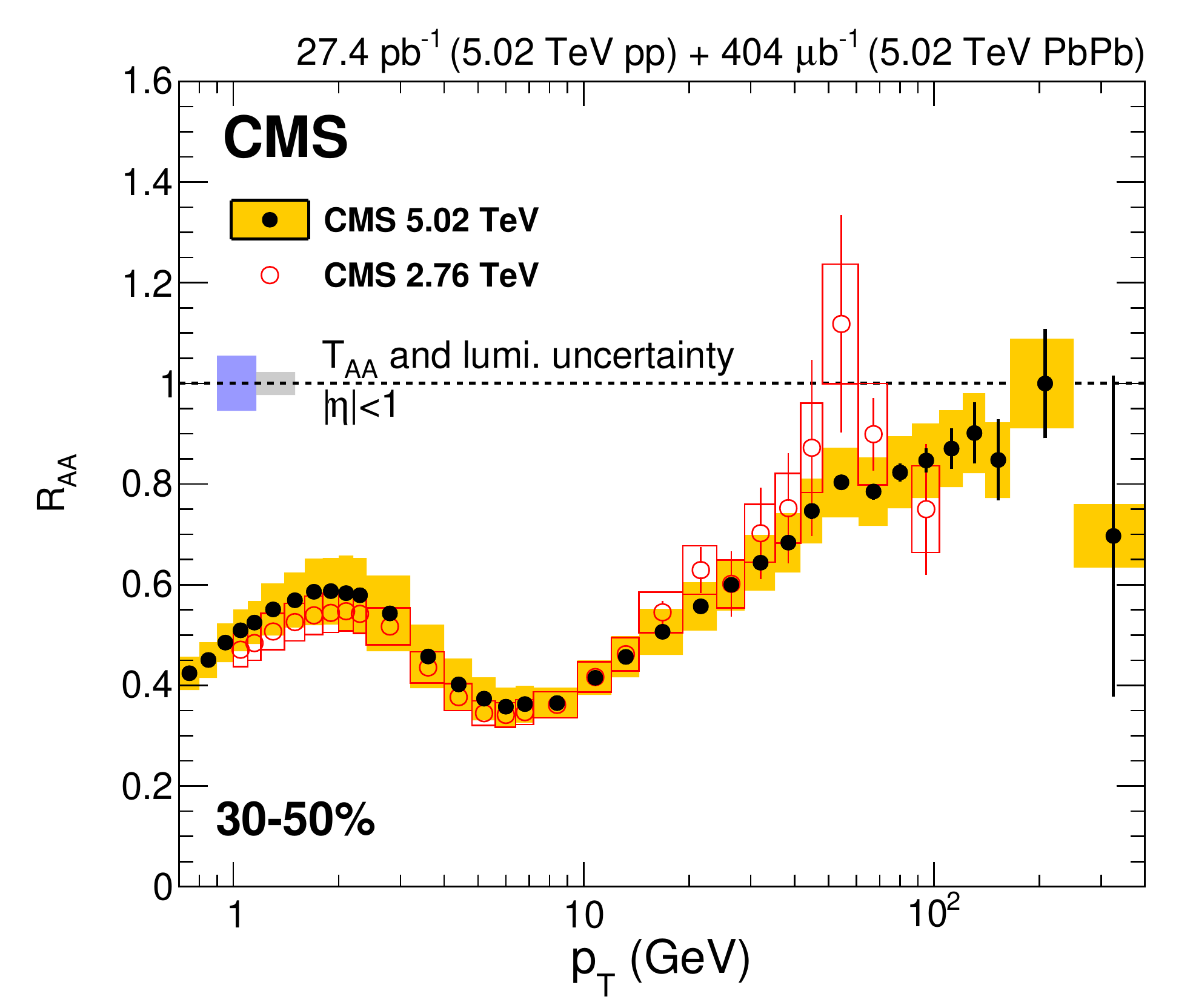}
    \includegraphics[width=0.45\textwidth]{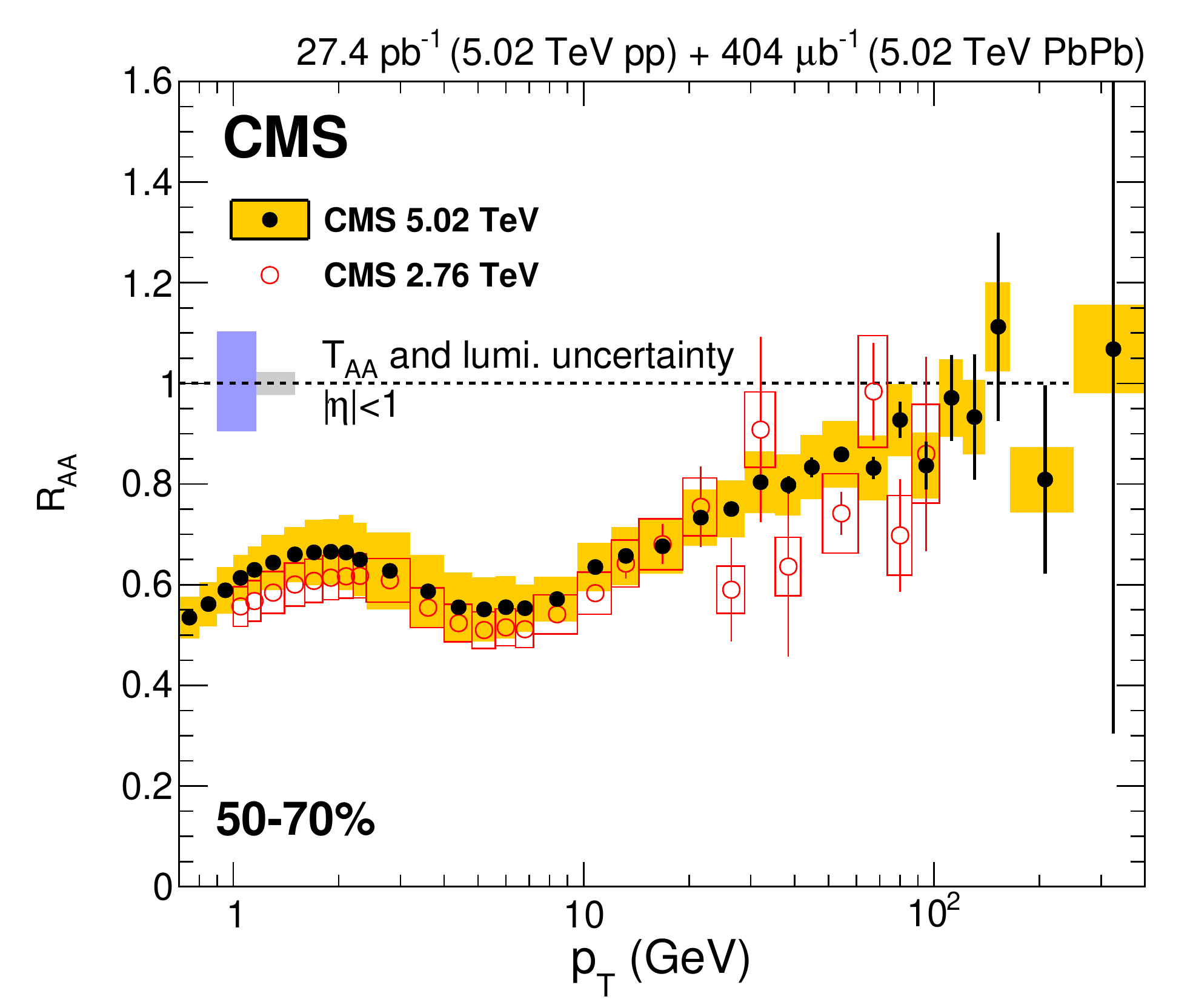}
    \includegraphics[width=0.45\textwidth]{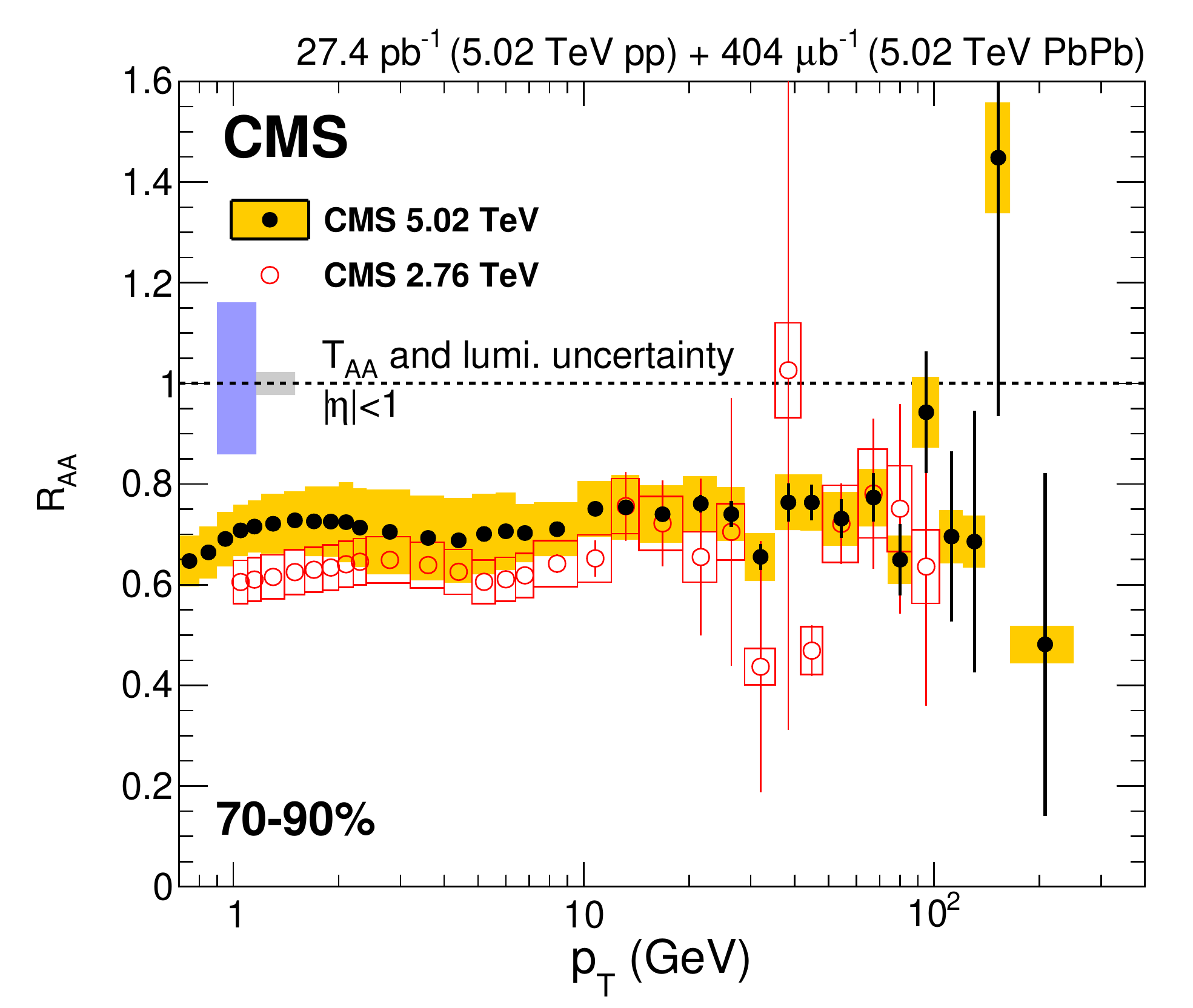}
  \caption{Charged-particle \raa measured in six different centrality
ranges at $\sqrtsnn=5.02$\TeV compared to
results at $\sqrtsnn=2.76$\TeV from CMS~\cite{CMS:2012aa}
(all centrality bins), ALICE~\cite{Abelev:2012hxa} (in the 0--5\% and
5--10\% centrality ranges), and ATLAS~\cite{Aad:2015wga} (in the 0--5\%
centrality range).  The yellow boxes represents the systematic
uncertainty of the 5.02\TeV CMS points.}
  \label{fig:RAA_Result_Compare276}
\end{figure*}

The measured charged-particle spectra are shown in Fig.~\ref{fig:SpectraResult}
for both pp and PbPb collisions at $\sqrtsnn=5.02$\TeV. The
PbPb results are shown in the 0--5\%, 5--10\%, 10--30\%, 30--50\%,
50--70\%, and 70--90\% centrality ranges, and are given as per-event
differential yields. The two most
central bins have been scaled by constant factors of three and ten for visual
clarity. The pp spectrum, for the purposes of measuring the \raa, is
measured as a differential cross section. In order to convert this
quantity to a per-event yield for comparison on the same figure, a scaling factor
of 70\unit{mb}, corresponding approximately to the total inelastic pp cross
section, is applied. No correction is applied for the finite size of the
\pt bins; the points represent the average yield across the bin. The
spectrum in pp collisions resembles a power law beyond a \pt of around
5\GeV. In comparison, the spectra in central PbPb collisions are visibly modified,
leading to \pt-dependent structures in \raa.
Representative systematic uncertainties are shown in the lower panel for central
and peripheral PbPb data, as well as for the pp data. The pp uncertainty shown
includes a 2.3\% correlated uncertainty coming from the use of the pp integrated
luminosity in the determination of the spectrum normalization.

The measured nuclear modification factors for primary charged particles
in PbPb collisions are shown in Fig.~\ref{fig:RAA_Result_Compare276}.
The error bars represent statistical uncertainties. The blue and gray
boxes around unity show the $T_\mathrm{AA}$ and pp
luminosity uncertainties, respectively, while the yellow band represents
the other systematic uncertainties as discussed in Section~\ref{sec:systUnc}.
The \raa distributions show a characteristic suppression pattern over
most of the \pt range measured, having local maxima at about a \pt of
2\GeV and local minima at around 7\GeV. These features are much stronger
for central collisions than for peripheral ones, and are presumably the
result of the competition between nuclear parton distribution function effects~\cite{Arneodo1994301}, radial flow~\cite{PhysRevC.87.014902}, parton energy loss, and
the Cronin effect~\cite{CroninEff,PhysRevC.88.024906f}, which all depend upon centrality. The suppression seen for 0--5\%
collisions is about 7--8 for \pt of around
6--9\GeV.  Above these \pt values, radial flow is insignificant and the
shape of \raa is expected to be dominated by parton energy loss. At
larger \pt, \raa appears to exhibit a continuous rise up to the highest
\pt values measured, with \raa values approaching unity. On the
other hand, the \raa for the 70--90\% centrality class displays relatively little \pt dependence. It is approximately centered around 0.75, albeit with a large systematic uncertainty which is dominated by a 16.1\% contribution from the $T_\mathrm{AA}$ uncertainty.  In all centrality classes, the uncertainties show a characteristic increase
in the 2--10\GeV \pt region driven by the uncertainty due to the
particle composition, which is largest in that region (see
Section~\ref{sec:systUnc}).

The measured \raa distributions at $\sqrtsnn=5.02$\TeV are
also compared to the CMS measurements at $\sqrtsnn=2.76$\TeV
\cite{CMS:2012aa} in Fig.~\ref{fig:RAA_Result_Compare276}.
Additionally, for the 0--5\% and 5--10\% bins, results from one or both
of the ALICE~\cite{Abelev:2012hxa} and ATLAS~\cite{Aad:2015wga} collaborations are shown.
The error bars represent the statistical uncertainties, while the boxes
indicate all systematic uncertainties, other than the luminosity and
$T_\mathrm{AA}$ uncertainties, for both CMS measurements. The 2.76\TeV CMS
measurement has a 6\% pp luminosity uncertainty and a $T_\mathrm{AA}$ uncertainty, which
is similar to that for 5.02\TeV~\cite{CMS:2012aa}.
The measured \raa distributions at 2.76 and 5.02\TeV are quantitatively similar to
each other. At \pt values below about 7\GeV, the 5.02\TeV data tend to be
higher, however the difference is mostly covered by the systematic
uncertainties of the respective measurements. It is worth noting that
because of the different particle composition corrections
applied in pp and PbPb at 5.02\TeV, the \raa is shifted upward by 1 to
5\% in the \pt region of 1--14\GeV compared to an \raa,
where no such correction is applied, such as the 2.76\TeV CMS
result. Above about
10\GeV and for central collisions, the 5.02\TeV \raa tends to be
slightly smaller than the 2.76\TeV one. For peripheral collisions, we
see the opposite trend.

\begin{figure}[t]
  \centering
    \includegraphics[width=0.45\textwidth]{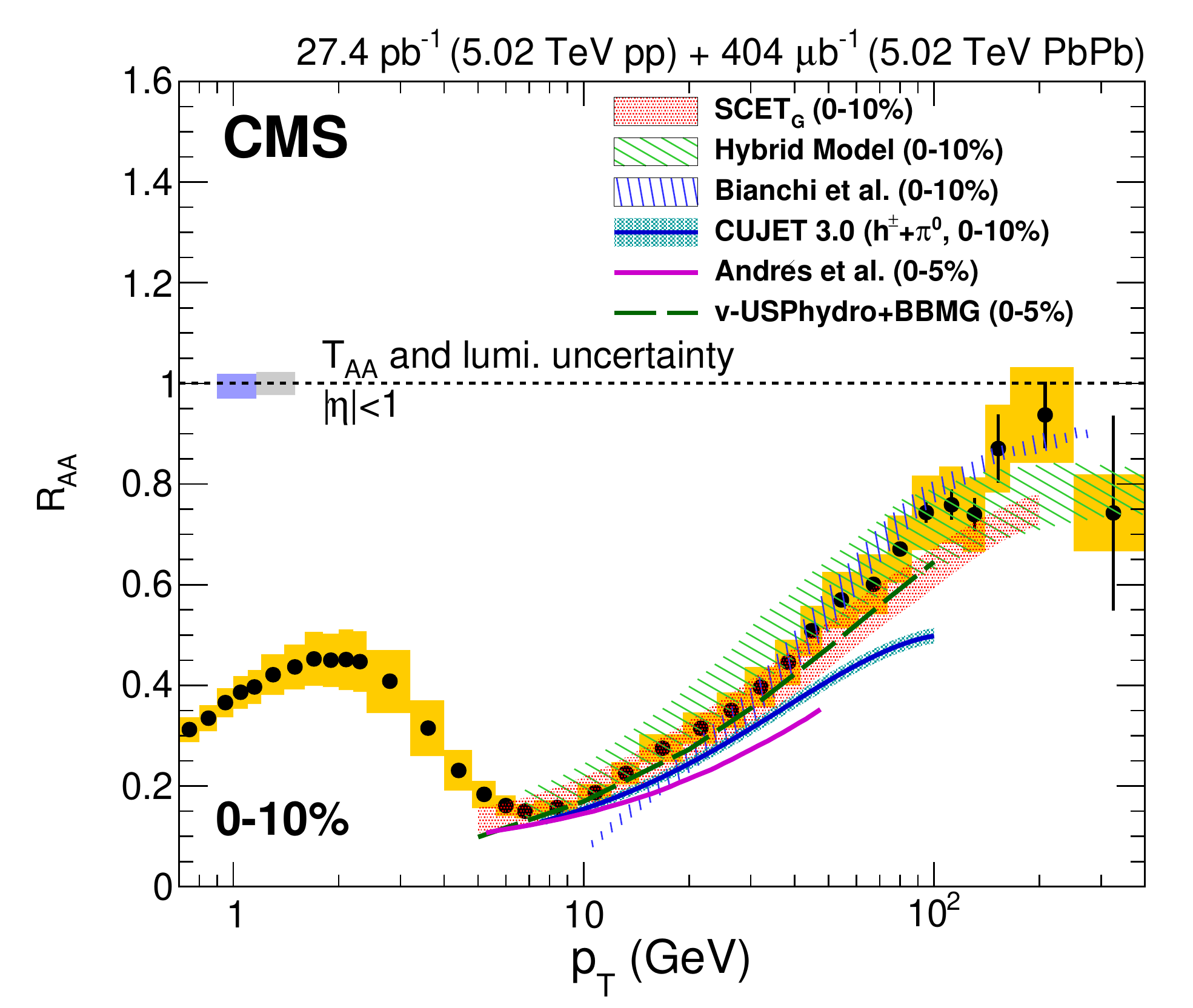}
    \includegraphics[width=0.45\textwidth]{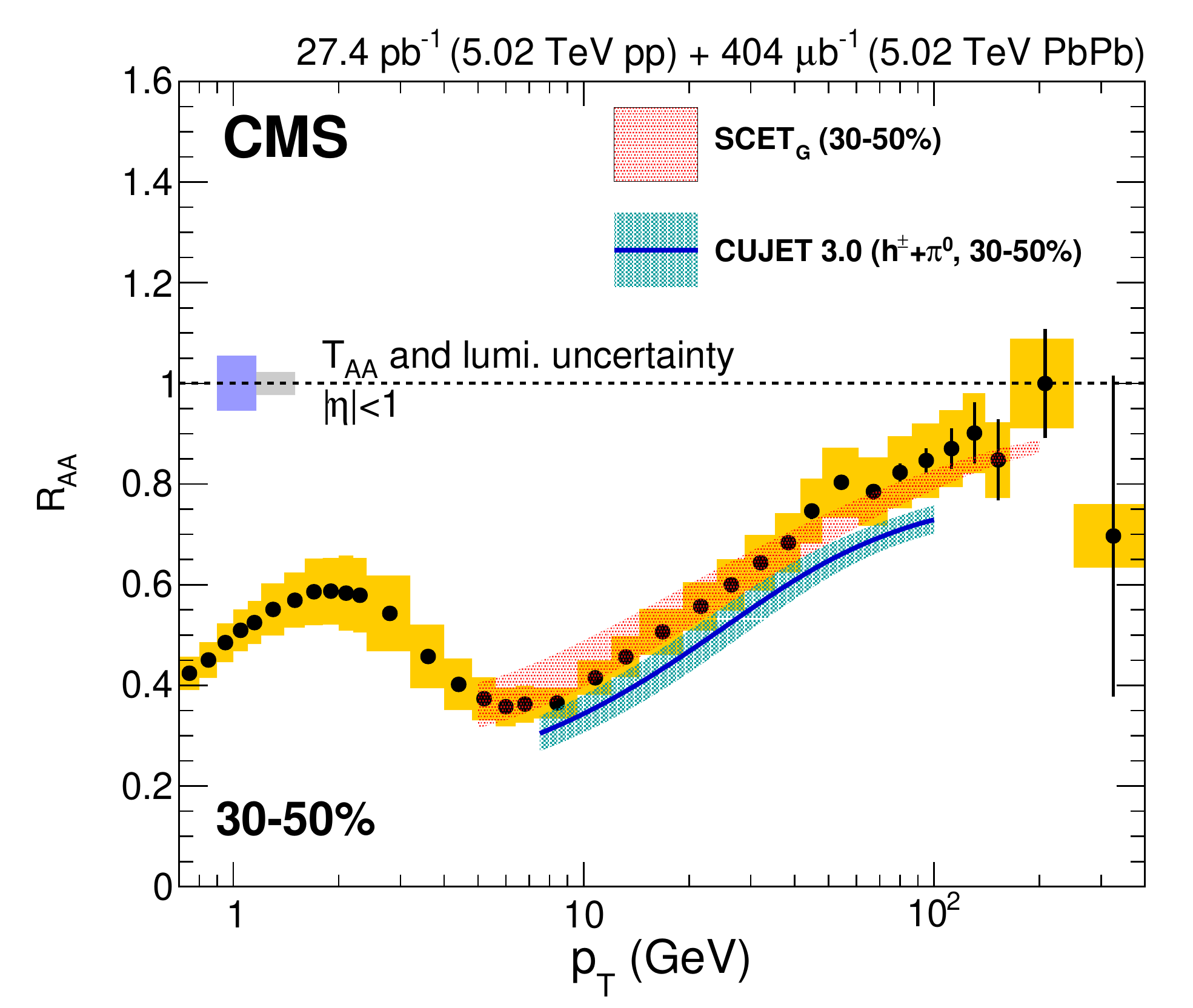}
  \caption{Charged-particle \raa measured in the 0--10\% ({left}) and
30--50\% ({right}) centrality ranges at $\sqrtsnn=5.02$\TeV compared
to predictions of models from
Refs.~\cite{Chien:2015vja,Hybrid_Model,jetscape,Xu:2015bbz,Andres:2016iys,Noronha-Hostler:2016eow}.
The yellow band represents the systematic uncertainty of the 5.02\TeV
CMS points.}
  \label{fig:RAA_Result_CompareTheory}
\end{figure}

Figure~\ref{fig:RAA_Result_CompareTheory} shows a comparison of the
measured \raa distributions in the 0--10\% and 30--50\% centrality ranges
to the predictions from models described in
Refs.~\cite{Chien:2015vja,Hybrid_Model,jetscape,Xu:2015bbz,Andres:2016iys,Noronha-Hostler:2016eow}.
The \textsc{scet}$_\textsc{G}$ model~\cite{Chien:2015vja} is based on the
generalization of the \textsc{dglap} evolution equations to include
final-state medium-induced parton showers combined with initial-state
effects. This model gives a good description of the measured data over
the full \pt range of the prediction, for \pt between 5
and 200\GeV. In the Hybrid model~\cite{Hybrid_Model}, the in-medium rate of energy loss is predicted using a strongly coupled theory.  This parametrization is then used to retroactively modify the particle shower produced by \PYTHIA~8.183.  Hadronization is accomplished using the \PYTHIA implementation of the Lund string model~\cite{HybridModel2}. The model tends to predict less suppression than the other models considered here, but is consistent with the measured data. The model of Bianchi et al.~\cite{jetscape} attempts to use the scale-dependence of the QGP parton distribution function to describe data at both RHIC and the LHC.  The calculation allows the medium transport coefficient, $\hat{q}$, to vary with the energy scale of jets traversing the medium.  Although the model agrees with the data well at high \pt, some discrepancy can be seen at the lower \pt range of the prediction.  The \textsc{cujet} 3.0 model~\cite{Xu:2015bbz} is constructed
by generalizing the perturbative-QCD-based \textsc{cujet} 2.0 model built
upon the Gyulassy--Levai--Vitev opacity series
formalism~\cite{Gyulassy:2001nm}. These generalizations
include two complementary nonperturbative features of the QCD
confinement cross-over phase transition: suppression of quark and gluon
degrees of freedom, and the emergence of chromomagnetic monopoles. For central collisions, the model predicts a suppression for charged hadrons plus neutral pions that is larger than seen in the data for charged particles.  In the 30--50\% centrality bin, however, the model is compatible with most of the data points.  The prediction by Andr\'es et al.~\cite{Andres:2016iys} comes from using the
'quenching weights' formalism and fitting a $K$ factor to the inclusive
particle suppression at LHC energies to parametrize the departure of $\hat{q}$ from an ideal estimate. The $K$
factor used to determine the predicted suppression at 5.02\TeV is
assumed to be the same as the one extracted from the fit to the 2.76\TeV
data. The predicted \raa shows a stronger suppression than the one seen
in data. As the authors note in Ref.~\cite{Andres:2016iys}, a $K$ value
needed to reproduce the CMS data is about 10\% smaller than the one
used. This indicates that the medium created at the higher collision
energy is closer to the ideal limit,
$\hat{q}\simeq2\varepsilon^{3/4}$~\cite{Baier:2002tc}, where
$\varepsilon$ is the energy density of the QGP.
Finally, the \textsc{v-usphydro+BBMG} model~\cite{Noronha-Hostler:2016eow}
couples event-by-event hydrodynamic flow and energy density profiles
calculated with \textsc{v-usphydro}~\cite{Noronha-Hostler:2013gga} to the
\textsc{BBMG} jet-energy-loss framework~\cite{Betz:2011tu}. For the curve
shown in Fig.~\ref{fig:RAA_Result_CompareTheory}, it is assumed that
the jet energy loss is proportional to the distance travelled in the
medium, that the shear viscosity to entropy density ratio of the medium is 0.05 (less than the Kovtun-Son-Starinets boundary of 1/4$\pi$~\cite{KSSLimit}), and that the freeze-out temperature is 160\MeV. The predicted \raa describes the data well lying on the lower
edge of the range covered by the systematic uncertainties of the
measurement.

\begin{figure}[tbh]
  \centering
    \includegraphics[width=\cmsFigWidth]{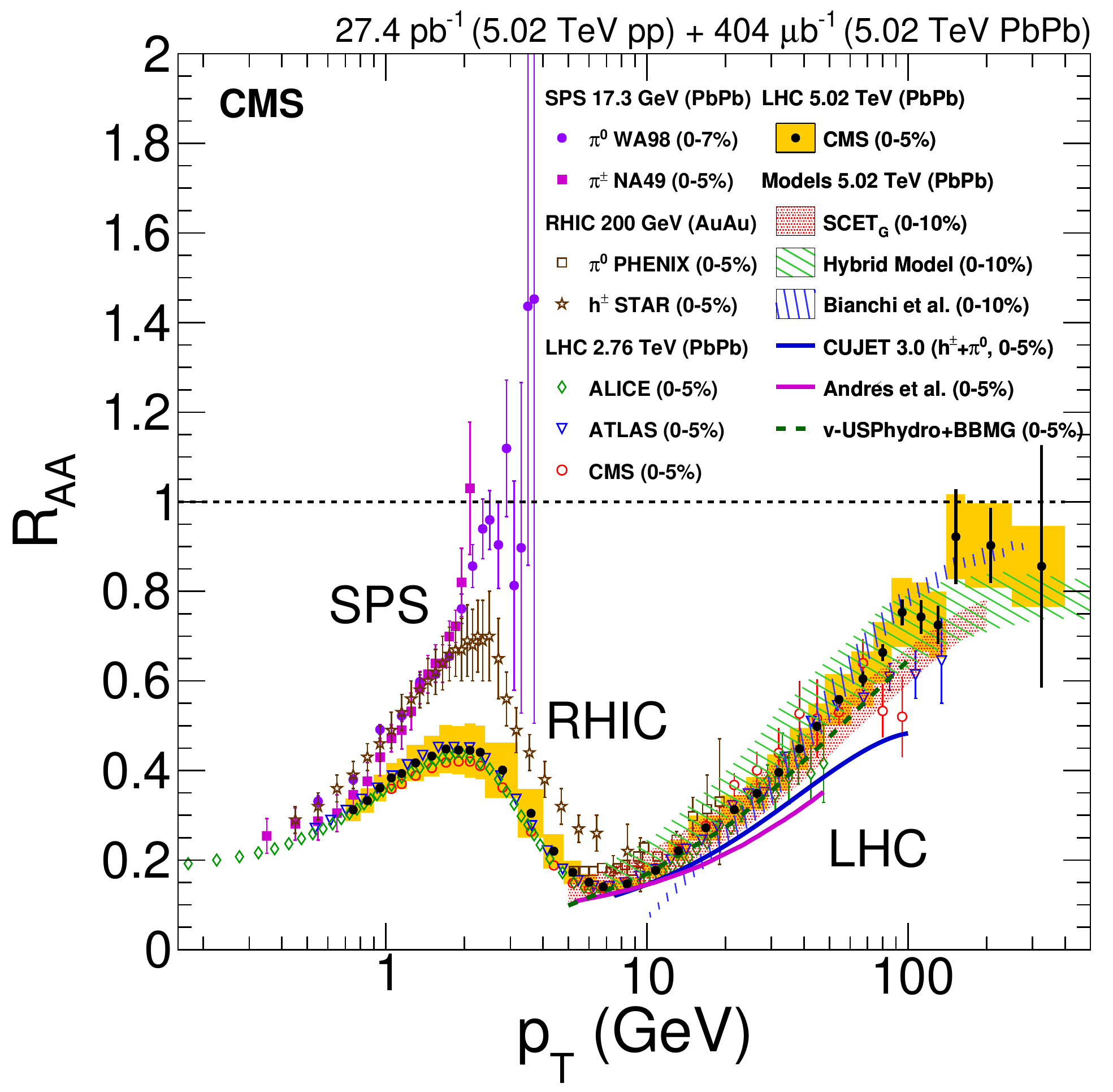}
  \caption{Measurements of the nuclear modification factors in central
heavy-ion collisions at four different center-of-mass energies, for
neutral pions (SPS, RHIC), charged hadrons ($h^{\pm}$) (SPS, RHIC), and charged
particles (LHC), from
Refs.~\cite{Aggarwal:2001gn,d'Enterria:2004ig,Alt:2007cd,Adare:2012wg,Adams:2003kv,Abelev:2012hxa,Aad:2015wga,CMS:2012aa},
compared to predictions of six models for $\sqrtsnn=5.02$\TeV
PbPb collisions from
Refs.~\cite{Chien:2015vja,Hybrid_Model,jetscape,Xu:2015bbz,Andres:2016iys,Noronha-Hostler:2016eow}.
The error bars represent the statistical uncertainties. The yellow band
around the 5.02\TeV CMS data points show the systematic uncertainties of
this measurement, including that of the integrated luminosity. The
$T_\mathrm{AA}$ uncertainties, of the
order of $\pm$5\%, are not shown. Percentage values in parentheses
indicate centrality ranges.}
  \label{fig:RAA_compilation}
\end{figure}

The evolution of central \raa with the collision center-of-mass energy, from the
SPS~\cite{Aggarwal:2001gn,d'Enterria:2004ig} to
RHIC~\cite{Adare:2012wg,Adams:2003kv}, and then to the
LHC~\cite{Abelev:2012hxa,Aad:2015wga,CMS:2012aa}, is presented in
Fig.~\ref{fig:RAA_compilation}. The data from WA98 and PHENIX are for
neutral pions, while the data given by NA49 and STAR are for charged
pions and hadrons, respectively.  The results from the present analysis
are shown by the black dots. The error bars show the statistical
uncertainties, while the yellow band surrounding the new $\sqrtsnn=5.02$\TeV CMS points represents the
systematic uncertainties, including that of the integrated luminosity (in the
previous figures the luminosity uncertainty is shown along with the
$T_\mathrm{AA}$ uncertainty as a separate error box around unity).
The $T_\mathrm{AA}$ uncertainties, which are less than 5\%, are not included
in the figure. The prediction of the models of
Refs.~\cite{Chien:2015vja,Hybrid_Model,jetscape,Xu:2015bbz,Andres:2016iys,Noronha-Hostler:2016eow}
at $\sqrtsnn=5.02$\TeV are also shown. The
measured nuclear modification factors at all energies show a rising
trend at low \pt up to 2\GeV, followed by local minima at RHIC and
the LHC at around 7\GeV. At higher \pt, both the RHIC and LHC data
show an increase of \raa with increasing \pt.

As the collision energy increases, high \pt charged-particle spectra flatten and extend to larger values.  If the average energy loss of a particle at a given \pt is fixed, this flattening would cause \raa to exhibit less suppression.  The similar \raa values measured at
2.76 and 5.02\TeV indicate that the effect of flattening spectra could be balanced by a larger average energy loss
in the higher-energy collisions at a fixed \pt~\cite{d'Enterria:2009am}.  A similar argument could explain the relatively close proximity of the 200\GeV PHENIX and 5.02\TeV CMS measurements for particle \pt$>$10\GeV, despite the latter having 25 times the collision energy.

\begin{figure}[tbh]
  \centering
    \includegraphics[width=\cmsFigWidth]{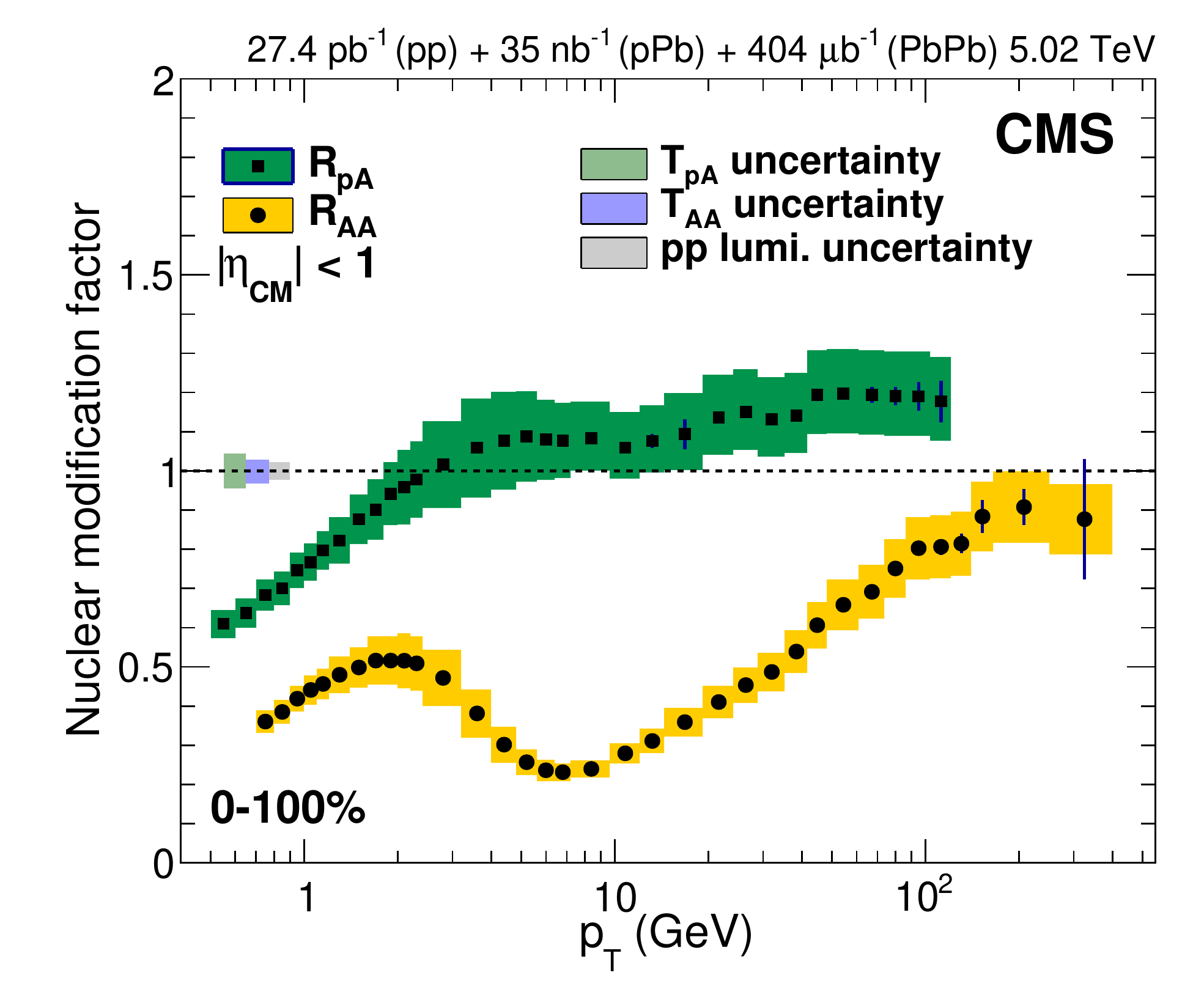}
  \caption{Measurements of the nuclear modification factor for an
inclusive centrality class for both PbPb and pPb collisions.  The
\rpa values are formed using the previously published CMS
pPb data~\cite{Khachatryan:2015xaa} and the pp reference spectrum
described in this paper. Please refer to the main text about the exact
procedure followed. The green and yellow boxes show the
systematic uncertainties for \rpa and \raa,
respectively, while the $T_{\Pp\mathrm{A}}$, $T_\mathrm{AA}$, and pp luminosity
uncertainties are shown as boxes at low \pt around unity.}
  \label{fig:RpPb}
\end{figure}

In order to better understand the relationship between the strong
suppression seen in \raa and potential cold nuclear matter effects, a
previous $\rpa^{*}$ measurement, using 35\nbinv of pPb data at
$\sqrtsnn=$5.02\TeV and an interpolated pp reference
\cite{Khachatryan:2015xaa}, is recalculated using the pp reference
spectrum measured in this paper at $\sqrt{s}=$5.02\TeV. In
order to do this, the corrections for the finite size of the \pt bins
applied to the published pPb data are
removed, as such a correction is not applied to the pp spectrum measured
here.  An additional correction for the particle species composition in
pPb collisions is calculated and applied in a fashion similar the
measured pp spectrum.  The previously published
data~\cite{Khachatryan:2015xaa} took this effect into account with a
systematic uncertainty, but the correction is applied here in order to
benefit from potential cancellations arising from the use of similar
analysis procedures on both spectra.  The systematic uncertainty due to
the particle composition effect was then updated in order to reflect the
presence of this additional correction.
Figure~\ref{fig:RpPb} shows the comparison between the nuclear
modification factors in inclusive pPb and PbPb collisions at
$\sqrtsnn=5.02$\TeV. At $\pt<2$\GeV a rising trend is
seen in both systems, which in PbPb collisions is followed by a
pronounced suppression in the $2<\pt<10$\GeV region, and a rising
trend from around 10\GeV to the highest \pt. In
the pPb system, there is no suppression in the intermediate \pt region, suggesting that in PbPb collisions the suppression is a hot
medium effect. Above $\pt>10$\GeV in the pPb system, a weak
momentum dependence is
seen leading to a moderate excess above unity at high \pt.  This excess is
less pronounced than the one seen in $\rpa^{*}$ when using an
interpolated pp reference spectrum~\cite{Khachatryan:2015xaa}.
At the \pt value of the largest deviation, 65\GeV, \rpa is $1.19 \pm0.02\stat{}^{+0.13}_{-0.11}\syst$, while $\rpa^{*}$ is $1.41\pm0.01\stat{}^{+0.20}_{-0.19}\syst$.
 The \rpa values above unity in the
intermediate \pt region are qualitatively similar to other observed
enhancements due to the Cronin effect and radial flow in pA and dA
systems~\cite{PhysRevC.88.024906f,Khachatryan:2015waa}. Furthermore,
the moderate excess above 10\GeV is suggestive of anti-shadowing
effects in the nuclear parton distribution
function~\cite{Arneodo1994301}.

\section{Summary}
\label{sec:summary}
The transverse momentum spectra of charged particles in pp and PbPb
collisions at $\sqrtsnn=5.02$\TeV have been measured in the pseudorapidity window
$\abs{\eta}<1$ in the \pt ranges of 0.5--400 (pp) and 0.7--400\GeV (PbPb). Using
these spectra, the nuclear modification factor \raa has been constructed in several bins
of collision centrality. In the 0--5\% bin, the \raa shows a maximum
suppression of a factor of 7--8 around \pt=7\GeV. At higher \pt, it exhibits a rise, reaching a value of $\raa=0.86\pm0.28$ in the \pt bin from 250 to 400\GeV. As collisions become more peripheral, a
weakening of both the magnitude and \pt dependence of this suppression
is observed.  Comparisons of the measured \raa values to the 2.76\TeV
results reveal similar \pt dependence and similar suppression.
Predictions of the high-\pt \raa coming from the \textsc{scet}$_\textsc{G}$, Hybrid, and \textsc{v-usphydro+BBMG} models
are found to approximately reproduce the present data. In central collisions, the
\textsc{cujet} 3.0 model and a model parametrizing the departure of the medium
transport coefficient, $\hat{q}$, from an ideal estimate, both
predict \raa suppressions that are
slightly larger than seen in data.  A model allowing $\hat{q}$ to vary is able to predict the data at high \pt, but expects a larger suppression around 10\GeV.  The nuclear modification factor in pPb collisions has been recomputed switching from an interpolation-based
reference to the newly measured pp data at $\sqrt{s}=5.02$\TeV. In the pPb system,
in contrast to the PbPb system, no suppression is observed in the 2--10\GeV region. A weak momentum dependence is seen for $\pt>
10$\GeV in the pPb system, leading to a moderate excess above unity at
high \pt.  The pPb and PbPb nuclear modification factors presented in this paper, covering \pt ranges up
to 120 and 400\GeV, respectively, provide stringent constraints on cold and hot nuclear matter effects.

\begin{acknowledgments}

We congratulate our colleagues in the CERN accelerator departments for the excellent performance of the LHC and thank the technical and
administrative staffs at CERN and at other CMS institutes for their contributions to the success of the CMS effort. In addition, we
gratefully acknowledge the computing centers and personnel of the
Worldwide LHC Computing Grid for delivering so effectively the computing
infrastructure essential to our analyses. Finally, we acknowledge the enduring support for the construction and operation of the LHC and the
CMS detector provided by the following funding agencies: BMWFW and FWF (Austria); FNRS and FWO (Belgium); CNPq, CAPES, FAPERJ, and FAPESP
(Brazil); MES (Bulgaria); CERN; CAS, MoST, and NSFC (China); COLCIENCIAS (Colombia); MSES and CSF (Croatia); RPF (Cyprus); SENESCYT
(Ecuador); MoER, ERC IUT and ERDF (Estonia); Academy of Finland, MEC, and HIP (Finland); CEA and CNRS/IN2P3 (France); BMBF, DFG, and HGF
(Germany); GSRT (Greece); OTKA and NIH (Hungary); DAE and DST (India); IPM (Iran); SFI (Ireland); INFN (Italy); MSIP and NRF (Republic of
Korea); LAS (Lithuania); MOE and UM (Malaysia); BUAP, CINVESTAV, CONACYT, LNS, SEP, and UASLP-FAI (Mexico); MBIE (New Zealand); PAEC
(Pakistan); MSHE and NSC (Poland); FCT (Portugal); JINR (Dubna); MON, RosAtom, RAS and RFBR (Russia); MESTD (Serbia); SEIDI and CPAN
(Spain); Swiss Funding Agencies (Switzerland); MST (Taipei); ThEPCenter, IPST, STAR and NSTDA (Thailand); TUBITAK and TAEK (Turkey); NASU
and SFFR (Ukraine); STFC (United Kingdom); DOE and NSF (USA).

Individuals have received support from the Marie-Curie programme and the European Research Council and EPLANET (European Union); the
Leventis Foundation; the A. P. Sloan Foundation; the Alexander von Humboldt Foundation; the Belgian Federal Science Policy Office; the Fonds
pour la Formation \`a la Recherche dans l'Industrie et dans l'Agriculture (FRIA-Belgium); the Agentschap voor Innovatie door Wetenschap en
Technologie (IWT-Belgium); the Ministry of Education, Youth and Sports (MEYS) of the Czech Republic; the Council of Science and Industrial
Research, India; the HOMING PLUS programme of the Foundation for Polish Science, cofinanced from European Union, Regional Development Fund,
the Mobility Plus programme of the Ministry of Science and Higher Education, the National Science Center (Poland), contracts Harmonia
2014/14/M/ST2/00428, Opus 2013/11/B/ST2/04202, 2014/13/B/ST2/02543 and 2014/15/B/ST2/03998, Sonata-bis 2012/07/E/ST2/01406; the Thalis and
Aristeia programmes cofinanced by EU-ESF and the Greek NSRF; the National Priorities Research Program by Qatar National Research Fund; the
Programa Clar\'in-COFUND del Principado de Asturias; the Rachadapisek Sompot Fund for Postdoctoral Fellowship, Chulalongkorn University and
the Chulalongkorn Academic into Its 2nd Century Project Advancement Project (Thailand); and the Welch Foundation, contract C-1845.

\end{acknowledgments}

\bibliography{auto_generated}

\providecommand{\href}[2]{#2}\begingroup\raggedright\begin{thebibliography}{10}%
\makeatletter
\providecommand{\hrefCMSnoop }[0]{\@secondoftwo}%
\makeatother
\providecommand{\doi}{\texttt{doi:}\begingroup \urlstyle{tt}\Url}

\bibitem{Bjorken:1982tu}
\href {http://lss.fnal.gov/archive/preprint/fermilab-pub-82-059-t.shtml}{J.~D.
  Bjorken, ``{Energy Loss of Energetic Partons in Quark-Gluon Plasma: Possible
  Extinction of High $p_{\rm T}$ Jets in Hadron-Hadron Collisions}'',}
  Technical Report {FERMILAB-PUB-82-059-T}, 1982.

\bibitem{d'Enterria:2009am}
\hrefCMSnoop {}{D.~d'Enterria, ``Jet quenching'',} in \textit{ Relativistic
  Heavy Ion Physics}, R.~Stock, ed., volume~23, p.~99.
\newblock 2010.
\newblock \href{http://www.arXiv.org/abs/0902.2011}{\texttt{arXiv:0902.2011}}.
\newblock {Landolt-B{\"o}rnstein/SpringerMaterials}.
\href{http://dx.doi.org/10.1007/978-3-642-01539-7_16}{\doi{10.1007/978-3-642-01539-7_16}}.

\bibitem{Aggarwal:2001gn}
\hrefCMSnoop {}{{WA98} Collaboration, ``{Transverse mass distributions of
  neutral pions from $^{208}$Pb-induced reactions at 158$\cdot A$~GeV}'',}
  \textit{ Eur. Phys. J. C} \textbf{ 23} (2002) 225,
  \href{http://dx.doi.org/10.1007/s100520100886}{\doi{10.1007/s100520100886}},
\href{http://www.arXiv.org/abs/nucl-ex/0108006}{\texttt{arXiv:nucl-ex/0108006}}.

\bibitem{d'Enterria:2004ig}
\hrefCMSnoop {}{D.~d'Enterria, ``{Indications of suppressed high $p_{\rm T}$
  hadron production in nucleus - nucleus collisions at CERN-SPS}'',} \textit{
  Phys. Lett. B} \textbf{ 596} (2004) 32,
  \href{http://dx.doi.org/10.1016/j.physletb.2004.06.071}{\doi{10.1016/j.physletb.2004.06.071}},
\href{http://www.arXiv.org/abs/nucl-ex/0403055}{\texttt{arXiv:nucl-ex/0403055}}.

\bibitem{Arsene:2004fa}
\hrefCMSnoop {}{{BRAHMS} Collaboration, ``Quark gluon plasma and color glass
  condensate at {RHIC}? The perspective from the {BRAHMS} experiment'',}
  \textit{ Nucl. Phys. A} \textbf{ 757} (2005) 1,
  \href{http://dx.doi.org/10.1016/j.nuclphysa.2005.02.130}{\doi{10.1016/j.nuclphysa.2005.02.130}},
\href{http://www.arXiv.org/abs/nucl-ex/0410020}{\texttt{arXiv:nucl-ex/0410020}}.

\bibitem{Back:2004je}
\hrefCMSnoop {}{{PHOBOS} Collaboration, ``The {PHOBOS} perspective on
  discoveries at {RHIC}'',} \textit{ Nucl. Phys. A} \textbf{ 757} (2005) 28,
  \href{http://dx.doi.org/10.1016/j.nuclphysa.2005.03.084}{\doi{10.1016/j.nuclphysa.2005.03.084}},
\href{http://www.arXiv.org/abs/nucl-ex/0410022}{\texttt{arXiv:nucl-ex/0410022}}.

\bibitem{Adams:2005dq}
\hrefCMSnoop {}{{STAR} Collaboration, ``{Experimental and theoretical
  challenges in the search for the quark gluon plasma: The STAR collaboration's
  critical assessment of the evidence from RHIC collisions}'',} \textit{ Nucl.
  Phys. A} \textbf{ 757} (2005) 102,
  \href{http://dx.doi.org/10.1016/j.nuclphysa.2005.03.085}{\doi{10.1016/j.nuclphysa.2005.03.085}},
\href{http://www.arXiv.org/abs/nucl-ex/0501009}{\texttt{arXiv:nucl-ex/0501009}}.

\bibitem{Adcox:2004mh}
\hrefCMSnoop {}{{PHENIX} Collaboration, ``{Formation of dense partonic matter
  in relativistic nucleus nucleus collisions at RHIC: Experimental evaluation
  by the PHENIX collaboration}'',} \textit{ Nucl. Phys. A} \textbf{ 757} (2005)
  184,
  \href{http://dx.doi.org/10.1016/j.nuclphysa.2005.03.086}{\doi{10.1016/j.nuclphysa.2005.03.086}},
\href{http://www.arXiv.org/abs/nucl-ex/0410003}{\texttt{arXiv:nucl-ex/0410003}}.

\bibitem{Abelev:2012hxa}
\hrefCMSnoop {}{{ALICE} Collaboration, ``Centrality dependence of charged
  particle production at large transverse momentum in {Pb--Pb} collisions at
  {$\sqrt{s_{\rm{NN}}} = 2.76$ TeV}'',} \textit{ Phys. Lett. B} \textbf{ 720}
  (2013) 52,
  \href{http://dx.doi.org/10.1016/j.physletb.2013.01.051}{\doi{10.1016/j.physletb.2013.01.051}},
\href{http://www.arXiv.org/abs/1208.2711}{\texttt{arXiv:1208.2711}}.

\bibitem{Aad:2015wga}
\hrefCMSnoop {}{{ATLAS} Collaboration, ``{Measurement of charged-particle
  spectra in Pb+Pb collisions at $\sqrt{{s}_\mathsf{{NN}}} = 2.76$ TeV with the
  ATLAS detector at the LHC}'',} \textit{ JHEP} \textbf{ 09} (2015) 050,
  \href{http://dx.doi.org/10.1007/JHEP09(2015)050}{\doi{10.1007/JHEP09(2015)050}},
\href{http://www.arXiv.org/abs/1504.04337}{\texttt{arXiv:1504.04337}}.

\bibitem{CMS:2012aa}
\hrefCMSnoop {}{{CMS} Collaboration, ``{Study of high-$p_{\rm T}$ charged
  particle suppression in PbPb compared to pp collisions at $\sqrt{s_{\rm
  NN}}=2.76$ TeV}'',} \textit{ Eur. Phys. J. C} \textbf{ 72} (2012) 1945,
  \href{http://dx.doi.org/10.1140/epjc/s10052-012-1945-x}{\doi{10.1140/epjc/s10052-012-1945-x}},
\href{http://www.arXiv.org/abs/1202.2554}{\texttt{arXiv:1202.2554}}.

\bibitem{Miller:2007ri}
\hrefCMSnoop {}{M.~L. Miller, K.~Reygers, S.~J. Sanders, and P.~Steinberg,
  ``{Glauber modeling in high energy nuclear collisions}'',} \textit{ Ann. Rev.
  Nucl. Part. Sci.} \textbf{ 57} (2007) 205,
  \href{http://dx.doi.org/10.1146/annurev.nucl.57.090506.123020}{\doi{10.1146/annurev.nucl.57.090506.123020}},
\href{http://www.arXiv.org/abs/nucl-ex/0701025}{\texttt{arXiv:nucl-ex/0701025}}.

\bibitem{Khachatryan:2015xaa}
\hrefCMSnoop {}{{CMS} Collaboration, ``Nuclear effects on the transverse
  momentum spectra of charged particles in {pPb} collisions at
  {$\sqrt{s_\mathrm{NN}}=5.02$\TeV}'',} \textit{ Eur. Phys. J. C} \textbf{ 75}
  (2015) 237,
  \href{http://dx.doi.org/10.1140/epjc/s10052-015-3435-4}{\doi{10.1140/epjc/s10052-015-3435-4}}.

\bibitem{Aad:2016zif}
\hrefCMSnoop {}{{ATLAS Collaboration}, ``{Transverse momentum, rapidity, and
  centrality dependence of inclusive charged-particle production in
  $\sqrt{s_{\rm NN}}=5.02$ TeV p+Pb collisions measured by the ATLAS
  experiment}'',} (2016).
  \href{http://www.arXiv.org/abs/1605.06436}{\texttt{arXiv:1605.06436}}.
Submitted to Phys. Lett. B.

\bibitem{ALICE:2012mj}
\hrefCMSnoop {}{{ALICE} Collaboration, ``Transverse Momentum Distribution and
  Nuclear Modification Factor of Charged Particles in {p-Pb} Collisions at
  {$\sqrt{s_{\rm NN}}=5.02$ TeV}'',} \textit{ Phys. Rev. Lett.} \textbf{ 110}
  (2013) 082302,
  \href{http://dx.doi.org/10.1103/PhysRevLett.110.082302}{\doi{10.1103/PhysRevLett.110.082302}},
\href{http://www.arXiv.org/abs/1210.4520}{\texttt{arXiv:1210.4520}}.

\bibitem{Chatrchyan:2008zzk}
\hrefCMSnoop {}{{CMS} Collaboration, ``The {CMS} experiment at the {CERN}
  {LHC}'',} \textit{ JINST} \textbf{ 3} (2008) S08004,
\href{http://dx.doi.org/10.1088/1748-0221/3/08/S08004}{\doi{10.1088/1748-0221/3/08/S08004}}.

\bibitem{Alver:2008zza}
\hrefCMSnoop {}{B.~Alver {et~al.}, ``Importance of correlations and
  fluctuations on the initial source eccentricity in high-energy
  nucleus-nucleus collisions'',} \textit{ Phys. Rev. C} \textbf{ 77} (2008)
  014906,
  \href{http://dx.doi.org/10.1103/PhysRevC.77.014906}{\doi{10.1103/PhysRevC.77.014906}},
\href{http://www.arXiv.org/abs/0711.3724}{\texttt{arXiv:0711.3724}}.

\bibitem{Chatrchyan:2011sx}
\hrefCMSnoop {}{{CMS} Collaboration, ``{Observation and studies of jet
  quenching in PbPb collisions at nucleon-nucleon center-of-mass energy = 2.76
  TeV}'',} \textit{ Phys. Rev. C} \textbf{ 84} (2011) 024906,
  \href{http://dx.doi.org/10.1103/PhysRevC.84.024906}{\doi{10.1103/PhysRevC.84.024906}},
\href{http://www.arXiv.org/abs/1102.1957}{\texttt{arXiv:1102.1957}}.

\bibitem{Agashe:2014kda}
\hrefCMSnoop {}{{Particle Data Group}, K.~A. Olive {et~al.}, ``{Review of
  Particle Physics}'',} \textit{ Chin. Phys. C} \textbf{ 38} (2014) 090001,
\href{http://dx.doi.org/10.1088/1674-1137/38/9/090001}{\doi{10.1088/1674-1137/38/9/090001}}.

\bibitem{DEVRIES1987495}
\hrefCMSnoop {}{H.~De~Vries, C.~W. De~Jager, and C.~De~Vries, ``Nuclear
  charge-density-distribution parameters from elastic electron scattering'',}
  \textit{ At. Data Nucl. Data Tables} \textbf{ 36} (1987) 495,
  \href{http://dx.doi.org/10.1016/0092-640X(87)90013-1}{\doi{10.1016/0092-640X(87)90013-1}}.

\bibitem{Cacciari:2008gp}
\hrefCMSnoop {}{M.~Cacciari, G.~P. Salam, and G.~Soyez, ``The anti-$k_t$ jet
  clustering algorithm'',} \textit{ JHEP} \textbf{ 04} (2008) 063,
  \href{http://dx.doi.org/10.1088/1126-6708/2008/04/063}{\doi{10.1088/1126-6708/2008/04/063}},
  \href{http://www.arXiv.org/abs/0802.1189}{\texttt{arXiv:0802.1189}}.

\bibitem{Cacciari:2011ma}
\hrefCMSnoop {}{M.~Cacciari, G.~P. Salam, and G.~Soyez, ``{FastJet user
  manual}'',} \textit{ Eur. Phys. J. C} \textbf{ 72} (2012) 1896,
  \href{http://dx.doi.org/10.1140/epjc/s10052-012-1896-2}{\doi{10.1140/epjc/s10052-012-1896-2}},
\href{http://www.arXiv.org/abs/1111.6097}{\texttt{arXiv:1111.6097}}.

\bibitem{Kodolova:2007hd}
\hrefCMSnoop {}{O.~Kodolova, I.~Vardanian, A.~Nikitenko, and A.~Oulianov,
  ``{The performance of the jet identification and reconstruction in heavy ions
  collisions with CMS detector}'',} \textit{ Eur. Phys. J. C} \textbf{ 50}
  (2007) 117,
\href{http://dx.doi.org/10.1140/epjc/s10052-007-0223-9}{\doi{10.1140/epjc/s10052-007-0223-9}}.

\bibitem{1748-0221-9-10-P10009}
\hrefCMSnoop {}{{CMS} Collaboration, ``{Description and performance of track
  and primary-vertex reconstruction with the CMS tracker}'',} \textit{ JINST}
  \textbf{ 9} (2014) P10009,
  \href{http://dx.doi.org/10.1088/1748-0221/9/10/P10009}{\doi{10.1088/1748-0221/9/10/P10009}},
\href{http://www.arXiv.org/abs/1405.6569}{\texttt{arXiv:1405.6569}}.

\bibitem{CMS-PAS-PFT-10-002}
\href {https://cds.cern.ch/record/1279341}{{CMS} Collaboration, ``Commissioning
  of the Particle-Flow reconstruction in Minimum-Bias and Jet Events from pp
  Collisions at 7 TeV'',} CMS Physics Analysis Summary CMS-PAS-PFT-10-002,
  2010.

\bibitem{Sjostrand:2007gs}
\hrefCMSnoop {}{T.~Sj{\"o}strand, S.~Mrenna, and P.~Z. Skands, ``A brief
  introduction to {PYTHIA 8.1}'',} \textit{ Comput. Phys. Commun.} \textbf{
  178} (2008) 852,
  \href{http://dx.doi.org/10.1016/j.cpc.2008.01.036}{\doi{10.1016/j.cpc.2008.01.036}},
\href{http://www.arXiv.org/abs/0710.3820}{\texttt{arXiv:0710.3820}}.

\bibitem{Khachatryan:2015pea}
\hrefCMSnoop {}{{CMS} Collaboration, ``{Event generator tunes obtained from
  underlying event and multiparton scattering measurements}'',} \textit{ Eur.
  Phys. J. C} \textbf{ 76} (2016) 155,
  \href{http://dx.doi.org/10.1140/epjc/s10052-016-3988-x}{\doi{10.1140/epjc/s10052-016-3988-x}},
\href{http://www.arXiv.org/abs/1512.00815}{\texttt{arXiv:1512.00815}}.

\bibitem{Lokhtin:2005px}
\hrefCMSnoop {}{I.~P. Lokhtin and A.~M. Snigirev, ``{A model of jet quenching
  in ultrarelativistic heavy ion collisions and high-$p_{\rm T}$ hadron spectra
  at RHIC}'',} \textit{ Eur. Phys. J. C} \textbf{ 45} (2006) 211,
  \href{http://dx.doi.org/10.1140/epjc/s2005-02426-3}{\doi{10.1140/epjc/s2005-02426-3}},
\href{http://www.arXiv.org/abs/hep-ph/0506189}{\texttt{arXiv:hep-ph/0506189}}.

\bibitem{MPT_Khachatryan:2015lha}
\hrefCMSnoop {}{{CMS} Collaboration, ``{Measurement of transverse momentum
  relative to dijet systems in PbPb and pp collisions at $
  \sqrt{s_{\mathrm{NN}}}=2.76 $ TeV}'',} \textit{ JHEP} \textbf{ 01} (2016)
  006,
  \href{http://dx.doi.org/10.1007/JHEP01(2016)006}{\doi{10.1007/JHEP01(2016)006}},
\href{http://www.arXiv.org/abs/1509.09029}{\texttt{arXiv:1509.09029}}.

\bibitem{Werner:2005jf}
\hrefCMSnoop {}{K.~Werner, F.-M. Liu, and T.~Pierog, ``{Parton ladder splitting
  and the rapidity dependence of transverse momentum spectra in deuteron-gold
  collisions at RHIC}'',} \textit{ Phys. Rev. C} \textbf{ 74} (2006) 044902,
  \href{http://dx.doi.org/10.1103/PhysRevC.74.044902}{\doi{10.1103/PhysRevC.74.044902}},
\href{http://www.arXiv.org/abs/hep-ph/0506232}{\texttt{arXiv:hep-ph/0506232}}.

\bibitem{Pierog:2013ria}
T.~Pierog\hrefCMSnoop {}{ {et~al.}, ``{EPOS LHC: Test of collective
  hadronization with data measured at the CERN Large Hadron Collider}'',}
  \textit{ Phys. Rev. C} \textbf{ 92} (2015) 034906,
  \href{http://dx.doi.org/10.1103/PhysRevC.92.034906}{\doi{10.1103/PhysRevC.92.034906}},
\href{http://www.arXiv.org/abs/1306.0121}{\texttt{arXiv:1306.0121}}.

\bibitem{ABELEV:2013zaa}
\hrefCMSnoop {}{{ALICE} Collaboration, ``{Multi-strange baryon production at
  mid-rapidity in Pb-Pb collisions at $\sqrt{s_{NN}}$ = 2.76 TeV}'',} \textit{
  Phys. Lett. B} \textbf{ 728} (2014) 216,
  \href{http://dx.doi.org/10.1016/j.physletb.2013.11.048}{\doi{10.1016/j.physletb.2013.11.048}},
  \href{http://www.arXiv.org/abs/1307.5543}{\texttt{arXiv:1307.5543}}.
[Corrigendum: \DOI{10.1016/j.physletb.2014.05.052}.

\bibitem{TRK-10-002}
\href {https://cds.cern.ch/record/1279139}{{CMS} Collaboration, ``Measurement
  of Tracking Efficiency'',} CMS Physics Analysis Summary CMS-PAS-TRK-10-002,
  2010.

\bibitem{Arneodo1994301}
\hrefCMSnoop {}{M.~Arneodo, ``Nuclear effects in structure functions'',}
  \textit{ Phys. Rep.} \textbf{ 240} (1994) 301,
  \href{http://dx.doi.org/10.1016/0370-1573(94)90048-5}{\doi{10.1016/0370-1573(94)90048-5}}.

\bibitem{PhysRevC.87.014902}
\hrefCMSnoop {}{{CMS} Collaboration, ``Measurement of the elliptic anisotropy
  of charged particles produced in PbPb collisions at $\sqrt{s_{\rm NN}}=2.76$
  TeV'',} \textit{ Phys. Rev. C} \textbf{ 87} (2013) 014902,
  \href{http://dx.doi.org/10.1103/PhysRevC.87.014902}{\doi{10.1103/PhysRevC.87.014902}},
  \href{http://www.arXiv.org/abs/1204.1409}{\texttt{arXiv:1204.1409}}.

\bibitem{CroninEff}
J.~W. Cronin\hrefCMSnoop {}{ {et~al.}, ``Production of hadrons at large
  transverse momentum at 200, 300, and 400 GeV'',} \textit{ Phys. Rev. D}
  \textbf{ 11} (1975) 3105,
  \href{http://dx.doi.org/10.1103/PhysRevD.11.3105}{\doi{10.1103/PhysRevD.11.3105}}.

\bibitem{PhysRevC.88.024906f}
\hrefCMSnoop {}{{PHENIX} Collaboration, ``Spectra and ratios of identified
  particles in {Au+Au} and {$d$+Au} collisions at {$\sqrt{s_{\rm NN}}=200$
  GeV}'',} \textit{ Phys. Rev. C} \textbf{ 88} (2013) 024906,
  \href{http://dx.doi.org/10.1103/PhysRevC.88.024906}{\doi{10.1103/PhysRevC.88.024906}},
  \href{http://www.arXiv.org/abs/1304.3410}{\texttt{arXiv:1304.3410}}.

\bibitem{Chien:2015vja}
Y.-T. Chien\hrefCMSnoop {}{ {et~al.}, ``Jet quenching from {QCD} evolution'',}
  \textit{ Phys. Rev. D} \textbf{ 93} (2016) 074030,
  \href{http://dx.doi.org/10.1103/PhysRevD.93.074030}{\doi{10.1103/PhysRevD.93.074030}},
\href{http://www.arXiv.org/abs/1509.02936}{\texttt{arXiv:1509.02936}}.

\bibitem{Hybrid_Model}
J.~Casalderrey-Solana\hrefCMSnoop {}{ {et~al.}, ``A hybrid strong/weak coupling
  approach to jet quenching'',} \textit{ JHEP} \textbf{ 10} (2014) 019,
  \href{http://dx.doi.org/10.1007/JHEP10(2014)019}{\doi{10.1007/JHEP10(2014)019}},
  \href{http://www.arXiv.org/abs/1405.3864}{\texttt{arXiv:1405.3864}}.
[Erratum: \DOI{10.1007/JHEP09(2015)175}].

\bibitem{jetscape}
E.~Bianchi\hrefCMSnoop {}{ {et~al.}, ``{The $x$ and $Q^2$ dependence of
  $\hat{q}$, quasi-particles and the JET puzzle}'',} (2017).
\href{http://www.arXiv.org/abs/1702.00481}{\texttt{arXiv:1702.00481}}.

\bibitem{Xu:2015bbz}
\hrefCMSnoop {}{J.~Xu, J.~Liao, and M.~Gyulassy, ``Bridging soft-hard transport
  properties of quark-gluon plasmas with {CUJET3.0}'',} \textit{ JHEP} \textbf{
  02} (2016) 169,
  \href{http://dx.doi.org/10.1007/JHEP02(2016)169}{\doi{10.1007/JHEP02(2016)169}},
\href{http://www.arXiv.org/abs/1508.00552}{\texttt{arXiv:1508.00552}}.

\bibitem{Andres:2016iys}
C.~Andr\'es\hrefCMSnoop {}{ {et~al.}, ``{Energy versus centrality dependence of
  the jet quenching parameter $\hat{q}$ at RHIC and LHC: a new puzzle?}'',}
  \textit{ Eur. Phys. J. C} \textbf{ 76} (2016) 475,
  \href{http://dx.doi.org/10.1140/epjc/s10052-016-4320-5}{\doi{10.1140/epjc/s10052-016-4320-5}},
\href{http://www.arXiv.org/abs/1606.04837}{\texttt{arXiv:1606.04837}}.

\bibitem{Noronha-Hostler:2016eow}
\hrefCMSnoop {}{J.~Noronha-Hostler, B.~Betz, J.~Noronha, and M.~Gyulassy,
  ``Event-by-Event Hydrodynamics $+$ Jet Energy Loss: A Solution to the
  {$R_{\rm AA} \otimes v_{\rm 2}$ Puzzle}'',} \textit{ Phys. Rev. Lett.}
  \textbf{ 116} (2016) 252301,
  \href{http://dx.doi.org/10.1103/PhysRevLett.116.252301}{\doi{10.1103/PhysRevLett.116.252301}},
\href{http://www.arXiv.org/abs/1602.03788}{\texttt{arXiv:1602.03788}}.

\bibitem{HybridModel2}
J.~Casalderrey-Solana\hrefCMSnoop {}{ {et~al.}, ``Angular Structure of Jet
  Quenching Within a Hybrid Strong/Weak Coupling Model'',} (2016).
\href{http://www.arXiv.org/abs/1609.05842}{\texttt{arXiv:1609.05842}}.

\bibitem{Gyulassy:2001nm}
\hrefCMSnoop {}{M.~Gyulassy, P.~Levai, and I.~Vitev, ``{Jet tomography of Au+Au
  reactions including multigluon fluctuations}'',} \textit{ Phys. Lett. B}
  \textbf{ 538} (2002) 282,
  \href{http://dx.doi.org/10.1016/S0370-2693(02)01990-1}{\doi{10.1016/S0370-2693(02)01990-1}},
\href{http://www.arXiv.org/abs/nucl-th/0112071}{\texttt{arXiv:nucl-th/0112071}}.

\bibitem{Baier:2002tc}
\hrefCMSnoop {}{R.~Baier, ``{Jet quenching}'',} \textit{ Nucl. Phys. A}
  \textbf{ 715} (2003) 209,
  \href{http://dx.doi.org/10.1016/S0375-9474(02)01429-X}{\doi{10.1016/S0375-9474(02)01429-X}},
\href{http://www.arXiv.org/abs/hep-ph/0209038}{\texttt{arXiv:hep-ph/0209038}}.

\bibitem{Noronha-Hostler:2013gga}
J.~Noronha-Hostler\hrefCMSnoop {}{ {et~al.}, ``Bulk viscosity effects in
  event-by-event relativistic hydrodynamics'',} \textit{ Phys. Rev. C} \textbf{
  88} (2013) 044916,
  \href{http://dx.doi.org/10.1103/PhysRevC.88.044916}{\doi{10.1103/PhysRevC.88.044916}},
\href{http://www.arXiv.org/abs/1305.1981}{\texttt{arXiv:1305.1981}}.

\bibitem{Betz:2011tu}
\hrefCMSnoop {}{B.~Betz, M.~Gyulassy, and G.~Torrieri, ``Fourier harmonics of
  high-$p_{T}$ particles probing the fluctuating intitial condition geometries
  in heavy-ion collisions'',} \textit{ Phys. Rev. C} \textbf{ 84} (2011)
  024913,
  \href{http://dx.doi.org/10.1103/PhysRevC.84.024913}{\doi{10.1103/PhysRevC.84.024913}},
\href{http://www.arXiv.org/abs/1102.5416}{\texttt{arXiv:1102.5416}}.

\bibitem{KSSLimit}
\hrefCMSnoop {}{P.~Kovtun, D.~T. Son, and A.~O. Starinets, ``{Holography and
  hydrodynamics: Diffusion on stretched horizons}'',} \textit{ JHEP} \textbf{
  10} (2003) 064,
  \href{http://dx.doi.org/10.1088/1126-6708/2003/10/064}{\doi{10.1088/1126-6708/2003/10/064}},
\href{http://www.arXiv.org/abs/hep-th/0309213}{\texttt{arXiv:hep-th/0309213}}.

\bibitem{Adare:2012wg}
\hrefCMSnoop {}{{PHENIX} Collaboration, ``{Neutral pion production with respect
  to centrality and reaction plane in Au$+$Au collisions at $\sqrt{s_{\rm NN}}$
  = 200 GeV}'',} \textit{ Phys. Rev. C} \textbf{ 87} (2013) 034911,
  \href{http://dx.doi.org/10.1103/PhysRevC.87.034911}{\doi{10.1103/PhysRevC.87.034911}},
\href{http://www.arXiv.org/abs/1208.2254}{\texttt{arXiv:1208.2254}}.

\bibitem{Adams:2003kv}
\hrefCMSnoop {}{{STAR} Collaboration, ``Transverse-Momentum and
  Collision-Energy Dependence of High $p_{T}$ Hadron Suppression in Au+Au
  COLLISions at Ultrarelativistic Energies'',} \textit{ Phys. Rev. Lett.}
  \textbf{ 91} (2003) 172302,
  \href{http://dx.doi.org/10.1103/PhysRevLett.91.172302}{\doi{10.1103/PhysRevLett.91.172302}},
\href{http://www.arXiv.org/abs/nucl-ex/0305015}{\texttt{arXiv:nucl-ex/0305015}}.

\bibitem{Alt:2007cd}
\hrefCMSnoop {}{{NA49} Collaboration, ``{High transverse momentum hadron
  spectra at $\sqrt{s_{\rm NN}} = 17.3$~GeV, in Pb+Pb and p+p collisions}'',}
  \textit{ Phys. Rev. C} \textbf{ 77} (2008) 034906,
  \href{http://dx.doi.org/10.1103/PhysRevC.77.034906}{\doi{10.1103/PhysRevC.77.034906}},
\href{http://www.arXiv.org/abs/0711.0547}{\texttt{arXiv:0711.0547}}.

\bibitem{Khachatryan:2015waa}
\hrefCMSnoop {}{{CMS} Collaboration, ``Evidence for Collective Multiparticle
  Correlations in pPbCollisions'',} \textit{ Phys. Rev. Lett.} \textbf{ 115}
  (2015) 012301,
  \href{http://dx.doi.org/10.1103/PhysRevLett.115.012301}{\doi{10.1103/PhysRevLett.115.012301}},
\href{http://www.arXiv.org/abs/1502.05382}{\texttt{arXiv:1502.05382}}.

\end{thebibliography}\endgroup

\cleardoublepage \appendix\section{The CMS Collaboration \label{app:collab}}\begin{sloppypar}\hyphenpenalty=5000\widowpenalty=500\clubpenalty=5000\textbf{Yerevan Physics Institute,  Yerevan,  Armenia}\\*[0pt]
V.~Khachatryan, A.M.~Sirunyan, A.~Tumasyan
\vskip\cmsinstskip
\textbf{Institut f\"{u}r Hochenergiephysik,  Wien,  Austria}\\*[0pt]
W.~Adam, E.~Asilar, T.~Bergauer, J.~Brandstetter, E.~Brondolin, M.~Dragicevic, J.~Er\"{o}, M.~Flechl, M.~Friedl, R.~Fr\"{u}hwirth\cmsAuthorMark{1}, V.M.~Ghete, C.~Hartl, N.~H\"{o}rmann, J.~Hrubec, M.~Jeitler\cmsAuthorMark{1}, A.~K\"{o}nig, I.~Kr\"{a}tschmer, D.~Liko, T.~Matsushita, I.~Mikulec, D.~Rabady, N.~Rad, B.~Rahbaran, H.~Rohringer, J.~Schieck\cmsAuthorMark{1}, J.~Strauss, W.~Waltenberger, C.-E.~Wulz\cmsAuthorMark{1}
% ...
\vskip\cmsinstskip
% ...
1:~~Also at Vienna University of Technology, Vienna, Austria\\
% ...

\end{sloppypar}
\end{document}